\documentclass[journal=jpccck,manuscript=article]{achemso}

\usepackage[version=3]{mhchem} % Formula subscripts using \ce{}

\usepackage{amsmath}
\usepackage{epsf}
\usepackage{epsfig}
\usepackage[T1]{fontenc}
\usepackage[latin9]{inputenc}
\usepackage{array}
\usepackage{multirow}

\author{C. Ataca}
\altaffiliation{UNAM-Institute of Materials Science and
Nanotechnology, Bilkent University, Ankara 06800, Turkey}
\affiliation{Department of Physics, Bilkent University, Ankara
06800, Turkey}

\author{S. Ciraci}\email{ciraci@fen.bilkent.edu.tr}
\affiliation{UNAM-Institute of Materials Science and
Nanotechnology, Bilkent University, Ankara 06800, Turkey}
\affiliation{Department of Physics, Bilkent University, Ankara
06800, Turkey}

\title[MoS2]
  {Functionalization of Single Layer MoS$_2$ Honeycomb Structures}

\abbreviations{IR,NMR,UV}

\begin{tocentry}
\includegraphics{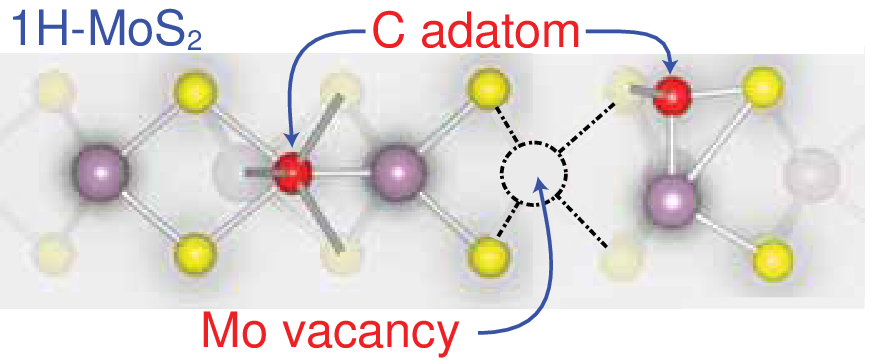}

\end{tocentry}

\date{\today}

\begin{document}

\begin{abstract}

Based on the first-principles plane wave calculations, we studied the functionalization of the two-dimensional single layer MoS$_2$ structure via adatom adsorption and vacancy defect creation. Minimum energy adsorption sites are determined for sixteen different adatoms, each gives rise to diverse properties. Bare, single layer MoS$_2$, which is normally a nonmagnetic, direct band gap semiconductor, attains a net magnetic moment upon adsorption of specific transition metal atoms, as well as silicon and germanium atoms. The localized donor and acceptor states in the band gap expand the utilization of MoS$_2$ in nanoelectronics and spintronics. Specific adatoms, like C and O, attain significant excess charge upon adsorption to single layer MoS$_{2}$ which may be useful for its tribological applications. Each MoS$_{2}$-triple vacancy created in a single layer MoS$_{2}$ gives rise to a net magnetic moment, while other vacancy defects related with Mo and S atoms do not influence the nonmagnetic ground state. Present results are also relevant for the surface of graphitic MoS$_2$.

\end{abstract}

\section{Introduction}

The synthesis of molybdenum disulfide, MoS$_2$, flakes have been subject of interest due to the unique symmetry of honeycomb structure and esoteric binding mechanism.\cite{Joenson,Helveg,Lee} Recently, 2D suspended single layer MoS$_2$ sheets (specified as 1H-MoS$_2$) have been produced.\cite{Joenson,novo2,Mak} Single layer MoS$_2$ nanocrystals of $\sim$ 30 \AA~ width were also synthesized on Au(111) surface.\cite{Helveg} Liquid exfoliation of nanosheets of MoS$_2$ and other transition metal dichalcogenides, such as MoSe$_2$, WS$_2$, MoTe$_2$, NbSe$_2$, NiTe$_2$, are reported.\cite{Coleman} 1H-MoS$_2$ is made of hexagons with Mo and two S atoms situated at alternating corners. Recently, this material with its nanoribbons has been an active field of theoretical studies.\cite{Li,Lebegue,Mendez,Bollinger,ataca,Galli}

MoS$_2$ has been used in ultra low friction studies. Formerly, the surface of 3D graphitic MoS$_2$ is used for these experiments to achieve a friction coefficient of $10^{-3}$ along the sliding direction in its basal planes.\cite{Martin} Recently, experiments\cite{Holscher, Lee2} using friction force microscopy and theoretical calculations\cite{Liang, Onodra} on atomically thin sheets of MoS$_2$ are confirmed the superlubricity of hexagonal MoS$_2$ structure. In addition to superlubricity, other areas, where MoS$_2$ appears to be a potential material for various technological applications, are hydrogen production,\cite{Hinnemann,Jaramillo} hydrodesulfurization catalyst used for removing sulfur compounds from oil,\cite{HDS2,HDS3,HDS4,HDS5,HDS6,HDS7,HDS8} solar cells,\cite{Kline} photocatalysis.\cite{Wilcoson} Photoluminescence is measured in monolayer MoS$_2$ which is absent in graphitic structure.\cite{splendiani} Most recently, a transistor fabricated from single layer MoS$_2$ has indicated the superior features of this 2D material over graphene.\cite{Radi} Studies to date suggests that MoS$_2$ can be promising for optoelectronic devices, solar cells, and LEDs.

Various properties revealed earlier for 3D graphitic MoS$_2$ and nowadays for 2D single layer 1H-MoS$_2$ have made the functionalization of these structures by adatoms or vacancy defects a current issue. Ab-initio calculations were carried for adsorption of thiophene\cite{Raybaud} on catalytically active surface of MoS$_2$ crystal, which is used in hydrodesulfurization process. Similarly, the adsorption of aromatic (thiophene, benzothiophene, benzene, naphthalene, pyridine, quinoline) and conjugated compounds (butadiene) on the basal plane (0001) of MoS$_2$ were studied.\cite{Moses} Huang and Cho\cite{Huang} investigated the adsorption of CO on MoS$_2$ surface and calculated the relative energies of different reaction paths. Implementing local magnetism through defects or impurities has been the focus of first-principles calculations. Fuhr \emph{et al.}\cite{Fuhr} found that S vacancy defect on MoS$_2$ surface, as well as substitutional doping of Pd and Au do not induce magnetic moment, whereas Fe and V induce magnetic moments when substituted with S atoms at MoS$_2$ surface. Adsorption and substitutional doping of Nb atoms on MoS$_2$ surface were also investigated.\cite{Ivanovskaya} Since the magnetism based on $sp$-orbitals yields long-range exchange coupling interactions as compared to $d-$ and $f-$ orbitals, implementing magnetic properties to MoS$_2$ monolayer through nonmagnetic adatoms have been also considered. Theoretical studies concluded that H, B, C, N and O can significantly modify the magnetic and electronic properties of this material.\cite{He}

In this paper we present a systematic study to reveal the general effects of the adsorption of selected adatoms and vacancy defect on the electronic and magnetic properties of perfect 1H-MoS$_2$. Three issues are of particular interest: (i) How can the surface charge of the single layer of MoS$_2$ be enhanced through adatoms? (ii) How can the spin polarization and magnetic moment be induced in nonmagnetic 1H-MoS$_2$? (iii) How can the electronic and magnetic states be modified through adatoms and vacancy induced localized gap states? We investigated 16 individual adatoms (such as C, Co, Cr, Fe, Ge, Mn, Mo, Ni, O, Pt, S, Sc, Si, Ti, V and W) and 5 different types of vacancy defects (such as Mo-, S-vacancy, S$_2$-, MoS-divacancy and MoS$_2$ triple vacancy) in view the above three issues and investigated their general features. Owing to the weak interlayer interaction present predictions concerning functionalization of 1H-MoS$_2$ honeycomb structure are expected to be relevant also for the MoS$_2$ sheets comprising a few layers, as well as the surface of graphitic MoS$_2$.

\section{Method}

Our results are based on first-principles plane wave calculations within density functional theory (DFT) using projector augmented wave (PAW) potentials.\cite{paw} The exchange correlation potential is approximated by generalized gradient approximation (GGA) using PW91\cite{pw91} functional both for spin-polarized and spin-unpolarized cases. While all discussions in the paper are based on the results obtained within GGA using PAW potential, calculations within local density approximation\cite{lda} (LDA) using PAW are also performed for specific cases for the purpose of comparison. All structures are treated within the periodic boundary conditions using supercell geometry. Kinetic energy cutoff, Brillouin zone (BZ) sampling are determined after extensive convergence analysis. A large spacing of $\sim$ 10 \AA~between 2D single layers of MoS$_2$ is taken to hinder the coupling between them. A plane-wave basis set with kinetic energy cutoff of 600 eV is used. In the self-consistent field potential and total energy calculations, BZ is sampled by special \textbf{k}-points.\cite{monk} The numbers of these \textbf{k}-points are (35x35x1) for the unitcell and (5x5x1) for the adatom adsorption in (4x4) supercell of 1H-MoS$_2$. All atomic positions and lattice constants are optimized by using the conjugate gradient method, where the total energy and atomic forces are minimized. The convergence for energy is chosen as 10$^{-5}$ eV between two consecutive steps, and the maximum Hellmann-Feynman forces acting on each atom is less than 0.02 eV/\AA~upon ionic relaxation. The pressure in the unit cell is kept below 1 kBar. Numerical calculations have been performed by using VASP.\cite{vasp1,vasp2}

Since DFT within GGA underestimates the band gap, frequency-dependent GW$_0$ calculations\cite{gw} are carried out to correct the band gaps. Screened Coulomb potential, W, is kept fixed to initial DFT value W$_{0}$ and Green's function, G, is iterated four times. Various tests regarding vacuum separation, kinetic energy cut-off energy, number of bands, \textbf{k}-points and grid points are made. Final results of GW$_{0}$ corrections are obtained using (12$\times$12$\times$1) \textbf{k}-points in BZ, 400 eV cut-off potential, 192 bands and 64 grid points for 1H-MoS$_2$. We were not able to apply GW$_{0}$ corrections for adatoms and vacancy due to large number of atoms in the supercells.

\begin{figure}
\centering
\includegraphics[width=8.5cm]{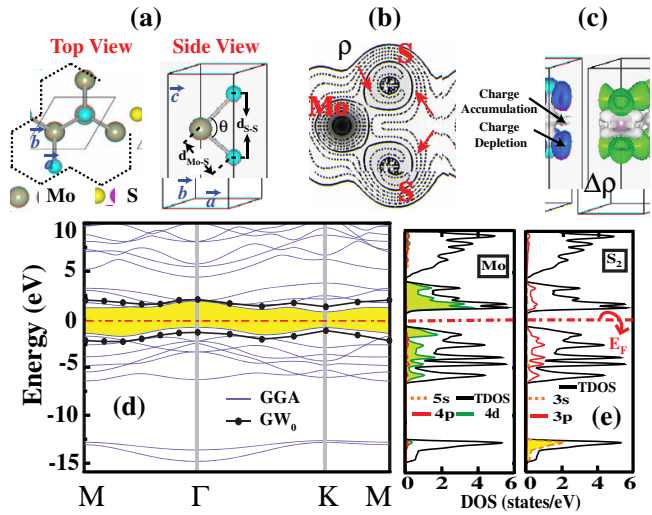}
\caption{(a) Top and side views of atomic structure of 2D 1H-MoS$_2$ with hexagonal lattice. The hexagonal unit cell with lattice constants $|a|=|b|$ is delineated by thin solid lines. Honeycomb structure consisting of Mo and S$_2$ atoms located at the corners is highlighted by dotted hexagons. (b) Contour plots of charge density, $\rho$ (see text for definition) in a vertical plane passing through Mo-S bonds. Arrows indicate the increasing value of charge density. (c) Isosurface plot of difference charge density, $\Delta \rho$ (see text for definition). Isosurface value is taken as 0.006 electrons/\AA$^{3}$. (d) Energy band structure of 1H-MoS$_2$ calculated by GGA+PAW using optimized structure. Zero of energy is set to the Fermi level indicated by dash-dotted line. The gap between valence and conduction band is shaded; GW$_0$ corrected valence and conduction bands are shown by filled circles. (e) Total density of states (TDOS) and orbital projected density of states (PDOS) for Mo and S.}
\label{fig:1}
\end{figure}

\section{Two dimensional honeycomb structure of MoS$_2$}

\begin{table*}
\caption{Calculated values of single layer 1H-MoS$_2$: Lattice constants, $|a|=|b|$; bond lengths, $d_{Mo-S}$ and $d_{S-S}$; S-Mo-S bond angle, $\theta$(S-Mo-S); cohesive energy per MoS$_2$ unit, $E_C$; direct energy band gap, $E_{g}$; band gap with GW$_0$ correction, $E_{g+GW_0}$; photoelectric threshold, $\Phi$.}
\label{tab:1h_mono}

\setlength{\extrarowheight}{1.5pt}
\tiny
\begin{tabular}{|c|c|c|c|c|c|c|c|c|}
\hline Method & $a$ (\AA) & $d_{Mo-S}$(\AA) & $d_{S-S}$(\AA) &
$\theta$(S-Mo-S) &$E_C$ (eV) & $E_{g} (eV)
$&$E_{g+GW_0}(eV)$&$\Phi$(eV)\tabularnewline \hline\
 GGA+PAW & 3.20
& 2.42 & 3.13 & 80.69$^{o}$ & 15.55 &1.58 &2.50 &5.96 \tabularnewline
\hline

LDA+PAW & 3.11 & 2.37 & 3.11 & 81.62$^{o}$ & 19.05  &1.87 &2.57 & -
\tabularnewline \hline

Experiment & 3.27 \cite{Yang}, 3.20 \cite{Joensen2}&  & - & - & -& -
& -&-\tabularnewline \hline

Theory & 3.13 \cite{Li}& 2.41 \cite{Li}& - & 82.31$^{o}$
\cite{Li}&15.6\cite{Li}, 14.99 \cite{Mendez} & 1.69 \cite{Li}, 1.8
\cite{Mendez}, 1.78 \cite{Lebegue}, 1.64 \cite{Bollinger} &
-&-\tabularnewline \hline
\end{tabular}
\end{table*}

To reveal the effects of adatoms and vacancy defects, we first present our results related with atomic, electronic and magnetic properties of 1H-MoS$_2$, which are obtained using the same calculation parameters. In ~\ref{fig:1} (a) we schematically illustrate the hexagonal crystal lattice, resulting in a honeycomb structure. The contour plots of total charge density, $\rho(\textbf{r}) = \sum^{occ} |\Psi(\textbf{k,r})|^{2} $, and difference charge density isosurfaces, $\Delta\rho (\textbf{r})$ (which is obtained by subtracting the free atom charge densities of Mo and S atoms from the total charge density of 1H-MoS$_2$) are also shown in ~\ref{fig:1} (b) and (c), respectively. There are ambiguities in determining the charge transfer; calculated excess charges may depend on the method used. While different methods result in different values for charge transfer (or excess charge on the atoms), its direction can be predicted unambiguously. Based on Mulliken\cite{Mulliken,Pauli} analysis the excess charge on each S atom and depletion of electrons on each Mo atom is calculated to be 0.205 electrons and 0.410 electrons, respectively. Bader analysis\cite{Bader} yields depletion of 1.2 electrons on Mo atom and 0.6 excess electrons on each S atom. Electronic charge transferred from Mo to S atoms gives rise to an excess charge around S atoms. Consequently, 1H-MoS$_2$ can be viewed as a positively charged Mo planes sandwiched between two negatively charged planes of S atoms as presented in \ref{fig:1} (c). The repulsive Coulomb force between negatively charged S atoms facing each other in adjacent MoS$_2$ layers weakens the interlayer interaction in two individual 1H-MoS$_2$ (or bilayer of MoS$_2$), as well as in graphitic MoS$_2$. In fact, the interaction between two adjacent MoS$_2$ layers of graphitic MoS$_2$ is repulsive even for large separation except a slight attractive range (of only 7 meV without van der Waals interaction, but $\sim$170 meV including van der Waals interaction) around the equilibrium separation. This is the main reason why MoS$_2$ flakes can be used to lower the friction coefficient in sliding friction. The cohesive energy of the optimized 1H-MoS$_2$ structure is found to be 15.55 eV per cell. For the known results, LDA+PAW calculations yield stronger binding and hence larger cohesive energy ($E_{C}$=19.05 eV).

Electronic structure of 2D suspended single layer 1H-MoS$_{2}$ and corresponding TDOS and PDOS calculated within GGA+PAW are presented in \ref{fig:1} (d) and (e). The states at the band edges, as well as at lower lying conduction and higher lying valence bands originate from the hybridization of Mo-$4d$ and S-$3p$ orbitals. The direct band gap is $E_{g}$=1.58 eV. The band gap calculated within LDA+PAW is $E_{g}$=1.87 eV, which is in good agreement with the band gap measured using complementary techniques of optical absorption, photoluminescence and photoconductivity as 1.90 eV.\cite{Mak} Earlier Bollinger \emph{et al.}\cite{Bollinger} predicted the band gap to be 1.64 eV using GGA+ultrasoft pseudopotential. However, Li and Galli\cite{Galli} with similar computational method calculated the band gap as 1.80 eV. Li \emph{et al.}\cite{Li} calculated the direct band gap 1.69 eV within GGA+PAW using relatively smaller energy cut-off. Mendez \emph{et al.}\cite{Mendez} found the direct band gap within LDA as $E_g$=1.8 eV using local basis set.\cite{siesta} Lebegue and Eriksson\cite{Lebegue} carried out LDA+PAW calculations using experimental lattice constants and found the band gap to be 1.78 eV. The band gap calculated in the present work is in fair agreement with previous studies.\cite{Bollinger,Li,Mendez} However, we show that band gaps determined in earlier studies increase $\sim$1 eV upon GW$_0$ correction. The band gap calculated within GGA(LDA)+PAW is corrected using self-energy method GW$_0$ to be 2.50(2.57) eV. The corrected band gap is $\sim$0.6-0.7 eV larger than the value measured experimentally.\cite{Mak} The situation with graphitic MoS$_2$, which consists of the stacking of 1H-MoS$_2$ layers is, however, different. The indirect band gap of graphitic MoS$_2$ calculated within GGA(LDA)+PAW is 0.85(0.72) eV and is corrected to be 1.44 (1.28) eV. In particular LDA+PAW band gap corrected by GW$_0$ is in good agreement with the experimental value.\cite{3dmos2} Since LDA/GGA is designed to describe systems with slowly varying electron density and may fail to model localized $d$-orbitals,\cite{Kresse} more accurate band gap calculations can be carried out using generalized Kohn-Sham scheme, screened nonlocal exchange functional, HSE.\cite{HSE} Starting from nonlocal charge density and wavefunctions, we calculated HSE and G$_0$W$_0$\cite{Kresse} corrected electronic band structure of MoS$_2$ as a direct band gap of 2.23 and 2.78 eV, respectively. Surprisingly, the band gap of 2D 1H-MoS$_2$ is overestimated by GW$_0$ correction. Here we note that the band gap of 2D fluorographene CF is also overestimated upon GW$_0$ correction.\cite{CFband} In \ref{tab:1h_mono} we list all the calculated structural parameters (including lattice constants, as well as internal parameters), cohesive energy, direct band gap together with GW$_0$ correction and photoelectric threshold of 2D 1H-MoS$_2$.

Finally, we emphasize two important dimensionality effects related with the electronic structure of MoS$_2$, which have important consequences like photoluminescence: (i) By going from graphitic MoS$_2$ to single layer 1H-MoS$_2$ the energy band gap changes from indirect to direct. (ii) The minimum gap increases by $\sim$0.6 eV. The transformation from indirect to direct gap is related with the orbital composition of states at the edges of conduction and valence bands.\cite{splendiani,Galli} The widening of the band gap in 2D occurs due to the absence of S-$p_z$ orbital interaction between adjacent MoS$_2$ layers and appears as the manifestation of the quantum confinement in the direction perpendicular to the MoS$_2$ layer.\cite{Mak}

\section{Functionalization by adatom adsorption}

\begin{figure*}
\centering
\includegraphics[width=14cm]{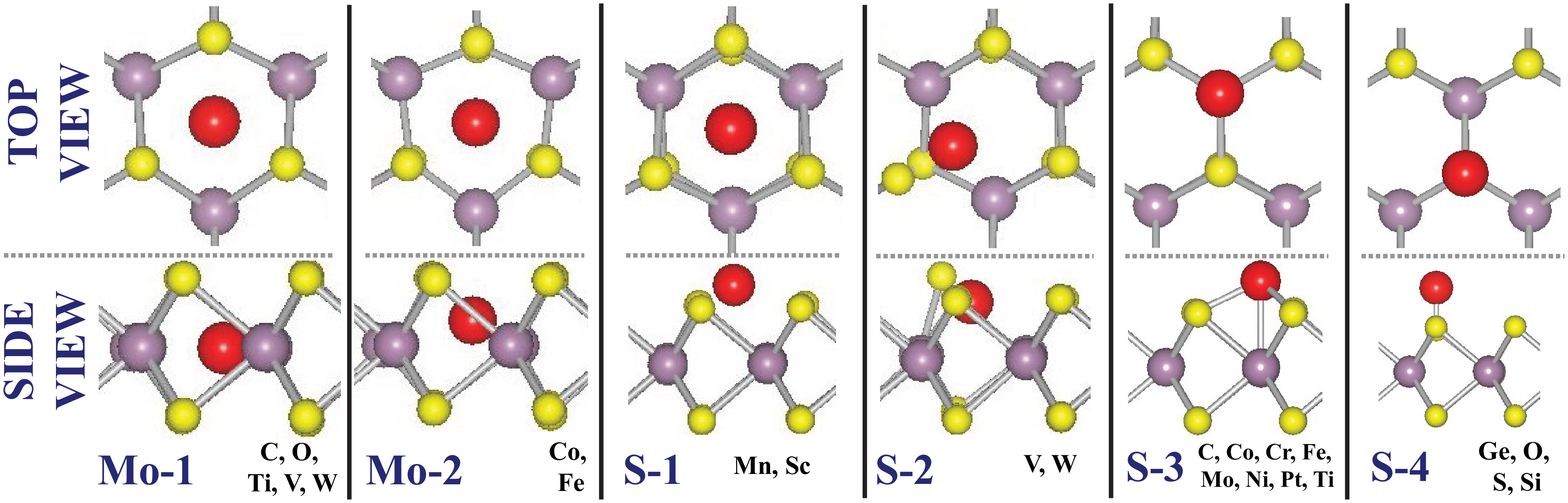}
\caption{ Top and side views are the schematic representation of possible adsorption geometries of adatoms obtained after the structure optimization. Adatoms, host Mo and S atoms are represented by red (large-dark), purple (medium-gray) and yellow (small-light) balls, respectively. Side views clarify the heights of adatoms from Mo and S atomic planes. Different adsorption sites are specified below each entry as 'Mo(S)-$\sharp$', where Mo(S) indicates that adatoms are placed initially (before structure optimization) to Mo(S) plane. In Mo-1 and Mo-2 geometries the adatoms are in and slightly above the Mo-layer. S-1, S-2,.. S-4 positions are associated with the S-layer. The adatoms adsorbed at each site are given at the lower right hand side of each entry. }
\label{fig:adatom-position}
\end{figure*}

Adsorption of adatoms is widely used and an efficient way to provide new functionalities to structures in nanoscale applications.\cite{canBN,canC,canSi,canLi,canCa,cohen,canNanowire} Among 16 different adatoms, namely, C, Co, Cr, Fe, Ge, Mn, Mo, Ni, O, Pt, S, Sc, Si, Ti, V and W, we were interested in 10 transition metal elements to determine ones which can induce magnetization in nonmagnetic 1H-MoS$_2$. Group 4A elements, i.e. C, Si and Ge are considered, since these atoms can form stable, planar or buckled honeycomb structures.\cite{novos,ansiklopedi,seymur} Carbon was of particular interest for the fabrication of graphene+MoS$_2$ complex or nanomeshes. Mo and S being host atoms in 1H-MoS$_2$, they can exist as residues. Whether MoS$_2$ can be oxidized is an important issue, which may limit future applications. The first question one has to address is, of course, whether these adatoms can form strong and stable bonding with 1H-MoS$_2$. The equilibrium adsorption sites of these 16 atoms are determined by first placing them to one of four different adsorption sites and subsequently by optimizing whole structure. Four possible adsorption sites considered initially for each adatom before the structure optimization are: (i) Hollow site slightly above the center of hexagon at Mo atomic plane. (ii) Hollow site above the center of hexagon at S-plane. (iii) Top of Mo atom. (iv) Top of S atom. Here the bridge site above the Mo-S bond is very similar to adsorption of adatom at the top of Mo atom. In order to avoid adatom-adatom coupling, the (4x4) supercell of 1H-MoS$_2$ is used, where the nearest adatom-adatom distance is $\sim 12.8$ \AA. Thus, the results can be related to the adsorption of an isolated adatom. The minimum energy positions of adatoms after the optimization process are described in \ref{fig:adatom-position}. We note that C, Co, Fe, O, Ti, V and W have two different binding sites (one having higher binding energy, $E_b$); each site leading to dramatically different electronic and magnetic structure. Six distinct adsorption sites described in \ref{fig:adatom-position}, namely Mo-1, Mo-2, S-1, S-2, S-3 and S-4 are distinguished after structure optimization.

All relevant data obtained from our calculations of adatoms adsorbed to 1H-MoS$_2$ are presented in \ref{tab:2da}. The height of the adatom from the Mo- or S-planes are calculated relative to the average heights of Mo- and S- atoms in the corresponding planes. The binding energy, $E_{b}$ is calculated as $E_{b} = E_{ad} + E_{MoS_2} - E_{ad+MoS_2}$. Here, $E_{ad}$ is the ground state energy of free adatom calculated in the same supercell with the same parameters; $E_{MoS_2}$ is the total energy of (4x4) supercell of 1H-MoS$_2$ and $E_{ad+MoS_2}$ is the optimized total energy of adatom+(4x4) supercell of 1H-MoS$_2$. Among different adatoms studied here, Cr has the weakest binding energy ($E_{b}=$1.08 eV); W has the strongest binding ($E_{b}=$4.93 eV) and creates a local reconstruction on S-layer. The excess charge\cite{Bader} of the adatom $\rho^{*}$, is obtained by subtracting the calculated charge at the adatom, $\rho_{A}$ from the valence charge of the adatom Z$_{A}$, namely $\rho^{*}=Z_{A}-\rho_{A}$. Accordingly $\rho^{*}$ < 0 implies excess electron at the adatom site. The magnetic moments are obtained by carrying out spin-polarized calculations without assigning initial atomic magnetic moments on adatoms. Upon relaxation the final magnetic moment is obtained.

\begin{table*}
\caption{Calculated values for the properties of 16 adatoms adsorbed on 1H-MoS$_2$. For specific adatoms, the first and second lines are associated with the adsorption to the Mo-layer and S-layer site, respectively. Other adatoms have only positive binding energy when adsorbed to the S-layer site. The adsorption sites of adatoms are described in \ref{fig:adatom-position}. $h_{Mo}$, the height of the adatom from Mo layer; $h_{S}$, the height of the adatom from the nearest S-layer; $d_{Mo}$, the adatom-nearest Mo distance; $d_{S}$, the adatom-nearest S distance; $E_b$, adatom binding energy; $\mu_{T}$, magnetic moment per supercell in Bohr magneton $\mu_{B}$; $\rho^{*}$, excess charge on the adatom (where negative sign indicates excess electrons); $\Phi$, photoelectric threshold (work function); P, dipole moment calculated in the direction normal to 1H-MoS$_2$ surface. $E_{i}$, energies of localized states induced by adatoms. Localized states are measured from the top of the valence bands in eV. The occupied ones are indicated by bold numerals and their spin alignments are denoted by either $\uparrow$ or $\downarrow$. States without the arrow sign indicating of spin alignment are nonmagnetic.}

\label{adatom}
\tiny
\begin{tabular}{ccccccccccccc}
\hline \hline \tabularnewline
  Atom & Site & $h_{Mo}$ &
$h_{S}$ & $d_{Mo}$ & $d_{S}$ &
 $E_{b}$  & $\mu_{T}$  & $\rho^{*}$ & $\Phi$ & P & E$_i$
 \tabularnewline

 & & (\AA) & (\AA) & (\AA) & (\AA) & (eV) & ($\mu_{B}$) & (e) & (eV) & (e
$\times$  \AA) & $\uparrow$ : Spin-up, $\downarrow$ : Spin-down States
\tabularnewline \hline \hline

\multirow{2}{*}{C} & Mo-1  & -0.01 & 1.56 & 2.04 & 2.44 & 3.28 & NM
&-0.65 & 5.81 & 0.00 & \textbf{0.14}, \textbf{0.19}, 2.18, 4.85
\tabularnewline

& S-3  & 1.58 & -0.05 & 2.07 & 1.80 & 2.69 & 2.00 & -0.58 & 5.74 &
-0.10 & \textbf{0.26} $\uparrow$,  \textbf{0.30} $\downarrow$, \textbf{1.07} $\uparrow$, \textbf{1.13} $\uparrow$, 1.55 $\downarrow$, 1.60 $\downarrow$
\tabularnewline

\hline%
 \multirow{2}{*}{Co} & Mo-2 & 0.61 & 0.99 & 2.22 & 2.19 & 0.96 & 1.00 & 0.15 & 5.57 & -0.07 & \textbf{0.61} $\uparrow$, \textbf{0.66} $\uparrow$, \textbf{0.82} $\downarrow$, \textbf{0.87} $\downarrow$, \textbf{0.89} $\uparrow$, 1.13 $\downarrow$
\tabularnewline

& S-3  & 2.52 & 0.94 & 2.56 & 2.11 & 2.92 & 1.00 & 0.44 & 5.17 &
-0.38 & \textbf{0.29} $\uparrow$, \textbf{0.51} $\uparrow$, \textbf{0.57} $\downarrow$, \textbf{0.62} $\uparrow$, \textbf{0.99} $\downarrow$, 1.00 $\downarrow$
\tabularnewline

\hline
 \multirow{1}{*}{Cr} & S-3  & 2.79 & 1.21 & 2.77 & 2.26 & 1.08 & 4.00 & 0.91 & 4.25 & -0.75 & \textbf{0.19} $\uparrow$, \textbf{0.22} $\uparrow$, \textbf{0.59} $\uparrow$,
  \textbf{0.64} $\uparrow$, \textbf{0.68} $\uparrow$, \textbf{1.60} $\downarrow$,  1.73 $\downarrow$
\tabularnewline

\hline
 \multirow{2}{*}{Fe} & Mo-2  & 0.31 & 1.29 & 2.21 & 2.26 & 0.39 & 2.00 & 0.34 & 5.07 & -0.04 & \textbf{0.32} $\uparrow$, \textbf{0.48} $\uparrow$, \textbf{0.87} $\uparrow$, \textbf{1.01} $\downarrow$, 1.15 $\downarrow$, 1.32 $\downarrow$
 \tabularnewline

& S-3  & 2.52 & 0.93 & 2.53 & 2.14 & 2.42 & 2.00 & 0.59 & 4.99 &
-0.46 & \textbf{-0.03} $\uparrow$, \textbf{0.18} $\uparrow$,
\textbf{0.21} $\uparrow$, \textbf{0.22} $\uparrow$, \textbf{0.91}
$\downarrow$, \textbf{0.93} $\downarrow$, 1.31 $\downarrow$
\tabularnewline

\hline
 \multirow{1}{*}{Ge} & S-4  & 3.83 & 2.26 & 4.28 & 2.30 & 1.18 & 2.00 & 0.39 & 4.78 & -0.27 & \textbf{1.10} $\uparrow$, \textbf{1.11} $\uparrow$, 1.72 $\downarrow$,  1.75 $\downarrow$
 \tabularnewline

\hline
 \multirow{1}{*}{Mn} & S-1  & 2.51 & 0.93 & 3.06 & 2.15 & 1.37 & 3.00 & 0.81 & 4.75 & -0.44 & \textbf{0.16} $\uparrow$, \textbf{0.17} $\uparrow$, \textbf{0.49} $\uparrow$, \textbf{0.50} $\uparrow$, \textbf{1.30} $\downarrow$,  1.98 $\downarrow$, 2.05 $\downarrow$
\tabularnewline

\hline
 \multirow{1}{*}{Mo} & S-3  & 2.89 & 1.31 & 2.84 & 2.30 & 1.43 & 4.00 & 0.81 & 4.54 & -0.76 & \textbf{0.48} $\uparrow$, \textbf{0.79} $\uparrow$, \textbf{0.87} $\uparrow$, \textbf{0.91} $\uparrow$, 1.16 $\downarrow$,  1.42 $\downarrow$, 1.54 $\downarrow$
\tabularnewline

\hline
 \multirow{1}{*}{Ni} & S-3  & 2.52 & 0.94 & 2.56 & 2.12 & 3.65 & NM & 0.36 & 5.58 & -0.31  & \textbf{0.12}, \textbf{0.16}, \textbf{0.21}, \textbf{0.50}, \textbf{0.51}, \textbf{0.57}
\tabularnewline

\hline
 \multirow{2}{*}{O} & Mo-1  & -0.01 & 1.57 & 1.95 & 2.56 & 2.24 & NM & -1.11 & 5.72 & 0.00 & \textbf{-6.36}, \textbf{-6.34}, \textbf{-5.66}, \textbf{0.32}
 \tabularnewline

& S-4 & 2.94 & 1.38 & 3.51 & 1.49 & 3.99 & NM & -0.91 & 5.96 & 0.35 & \textbf{-5.63}, \textbf{-1.16}, \textbf{-0.90}
\tabularnewline

\hline
 \multirow{1}{*}{Pt} & S-3  & 2.78 & 1.21 & 2.76 & 2.31 & 2.92 & NM & 0.08 & 5.66 & -0.26 & \textbf{0.26}, \textbf{0.27}, \textbf{0.37}, 1.76
\tabularnewline

\hline
 \multirow{1}{*}{S} & S-4  & 3.48 & 1.91 & 3.96 & 1.94 & 2.17 & NM & -0.11 & 5.96 & 0.23  & \textbf{0.00}, \textbf{0.06}, \textbf{0.11}
 \tabularnewline

\hline
 \multirow{1}{*}{Sc} & S-1  & 2.84 & 1.25 & 3.30 & 2.29 & 2.63 & 1.00 & 1.45 & 4.31 & -1.18 & \textbf{1.28} $\uparrow$, \textbf{1.31} $\uparrow$, \textbf{1.35} $\downarrow$, 1.49 $\downarrow$, 2.25 $\downarrow$, 2.39 $\uparrow$
\tabularnewline

\hline
 \multirow{1}{*}{Si} & S-4  & 3.73 & 2.16 & 4.18 & 2.17 & 1.39 & 2.00 & 0.58 & 4.94 & -0.14 & \textbf{0.98} $\uparrow$, \textbf{0.99} $\uparrow$, 1.65 $\downarrow$, 1.68 $\downarrow$
  \tabularnewline

\hline
 \multirow{2}{*}{Ti} & Mo-1  & 0.00 & 1.58 & 2.31 & 2.42 & 1.23 & NM & 1.14 & 5.71 & 0.00 & \textbf{0.26}, \textbf{0.34}, \textbf{0.36}, 1.74, 1.77
\tabularnewline

& S-3  & 2.95 & 1.38 & 2.99 & 2.32 & 2.62 & 4.00 &1.16 & 4.32 & -1.12 & \textbf{1.18} $\uparrow$, \textbf{1.24} $\uparrow$, \textbf{1.35} $\uparrow$, \textbf{1.37} $\uparrow$, 1.54 $\uparrow$, 1.77 $\downarrow$, 1.81 $\downarrow$
 \tabularnewline

\hline
 \multirow{2}{*}{V} & Mo-1  & 0.03 & 1.55 & 2.95 & 2.34 & 1.25 & 1.00& 1.05 & 4.61 & 0.00 & \textbf{0.12} $\uparrow$, \textbf{0.16} $\downarrow$, \textbf{0.17} $\uparrow$, \textbf{0.21} $\downarrow$,  \textbf{1.54} $\uparrow$, 1.67 $\uparrow$, 1.69 $\downarrow$
\tabularnewline

& S-2  & 1.87& 0.19 & 2.56 & 2.07& 2.76 & 1.00 & 1.05& 5.18 & 0.04 & \textbf{0.07} $\uparrow$, \textbf{0.18} $\downarrow$, \textbf{0.84} $\uparrow$, 1.23 $\downarrow$, 1.45 $\uparrow$, 1.54 $\downarrow$
\tabularnewline

\hline
 \multirow{2}{*}{W} & Mo-1  & 0.01 & 1.58 & 2.41 & 2.42 & 1.18 & 2.00 & 1.05 & 4.66 & 0.01 &  \textbf{0.16} $\uparrow$, \textbf{0.21} $\downarrow$, \textbf{1.41} $\uparrow$, \textbf{1.49} $\uparrow$, 1.61 $\downarrow$, 1.81 $\downarrow$
 \tabularnewline

& S-2  & 1.87& 0.16& 2.61 & 2.15& 4.93& NM& 0.85& 5.58& 0.12 &  \textbf{0.13}, \textbf{0.19}, \textbf{0.58}, 1.57, 1.75
 \tabularnewline

\hline \hline

\end{tabular}
\label{tab:2da}
\end{table*}

Since the adatom-adatom interaction is hindered by a large separation between them, the adatoms presented in \ref{tab:2da} give rise to localized electronic states in the band gap and resonant states in band continua and hence modify the electronic properties of 1H-MoS$_2$. In \ref{fig:adatom-band}, localized states of O, Ti, Cr and Ge together with band decomposed charge density isosurfaces are presented. These atoms are specifically selected, since they are representatives of some of the adatoms presented \ref{tab:2da}. Oxygen adatom when placed on the S plane is adsorbed on top of S atom (S-4 site) with a binding energy of 3.99 eV. This site is in agreement with the results of He \emph{et al.}\cite{He} However, we predict also a local minimum at Mo-1 site with relatively smaller binding energy ($E_{b}=$2.44 eV). Adsorbed O is nonmagnetic at both sites. Oxygen adatom having the highest electronegativity and highest negative excess charge among all other adatoms have localized states in the valence band. When O is adsorbed at Mo-l site, the only localized state occurring in the band gap is filled and originate from the combination of O-$p_z$ orbital with the $p$-orbitals of nearest S atoms. Sulfur being in the same group with O displays similar electronic properties and have localized states in the band gap just above the valence band originated from its $p_x$ and $p_y$ orbitals.

\begin{center}
\begin{figure*}
\includegraphics[width=14cm]{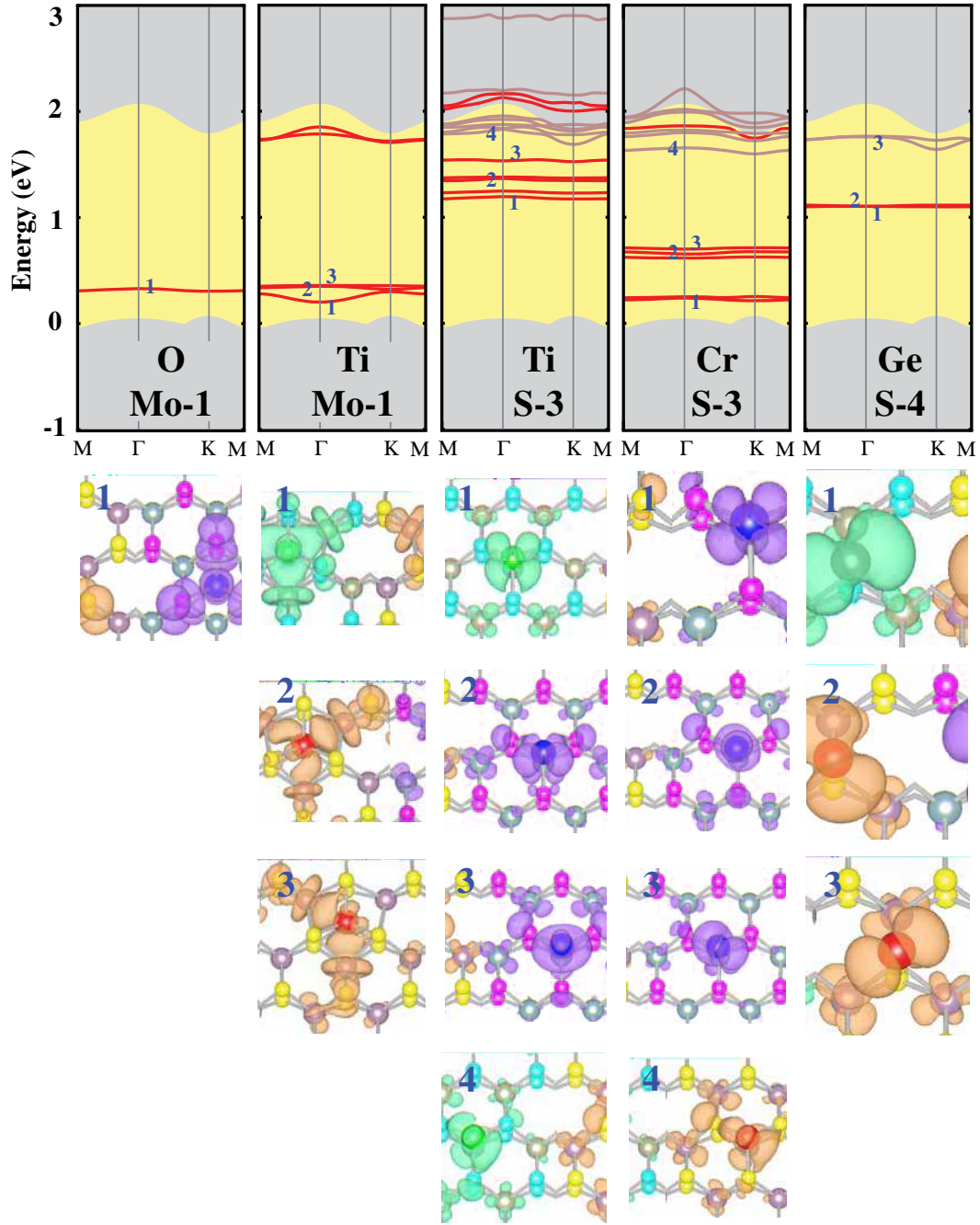}
\caption{Schematic diagram of the relevant energy levels (or bands) of single adatom (O, Ti, Cr and Ge) adsorbed to each (4$\times$4) supercell of 1H-MoS$_2$. The grey (light) shaded region in the background is the valence and conduction band continua. For nonmagnetic case, red (dark) bands are contributed more than 50\% by adatoms orbitals, For magnetic case, spin-up and spin-down bands are shown by red (dark) and brown (light) lines, respectively. Solid bands indicate that the contribution of adatom to the band is more than 50\%. In the lower part of each panel the adsorption site is indicated by the labeling of \ref{fig:adatom-position}. Charge density isosurfaces of adatom states specified by numerals are shown below. The isosurface value is taken as $2\times10^{-5}$ electrons/\AA$^3$.}
\label{fig:adatom-band}
\end{figure*}
\par\end{center}

Titanium can be adsorbed to both Mo and S layers. In \ref{fig:adatom-band} localized states of both adsorption geometries are indicated. When Ti adsorbed to Mo-l site, nearest $d_{z^2}$ orbitals of Mo atoms play an important role in the binding mechanism. Charge density isosurface of the state numbered as '1' in \ref{fig:adatom-band} is dominated from Ti's $d_{z^2}$, whereas isosurfaces '2' and '3' are dominated by $d_{xy}$ and $d_{x^2-y^2}$ of Ti which are parallel to Mo-layer. As for Ti adsorbed at S-3 site, Ti-$d_{xy}$ and $d_{xz}$ orbitals states dominate states numbered as '1' and '2' but the state numbered as '3' originates mainly from Ti-$d_{z^{2}}$ orbital. Cr and Mo adatoms at S-3 site on 1H-MoS$_2$ surface have four $d$-states singly occupied, but d$_{xy}$ state is doubly occupied. Cr-$s$ and Mo-$s$ orbitals are vacant in both adatoms. Hence Cr and Mo has local magnetic moment of $\mu=4 \mu_B$. As for W adatom at S-2 site, its localized magnetic moment of $\mu=2 \mu_B$ occurs from the spin-polarized $d$-orbitals of W together with the $p$-orbitals of nearest S-atom.

Carbon adatom is of particular interest. Previously, He \emph{et al.}\cite{He} found that the lowest energy adsorption position of C adatom occurs in the plane of sulfur atoms on top of Mo atom (namely S-3 site shown in \ref{fig:adatom-position}). Here, we found the same site with a binding energy of $E_b$=2.69 eV. However, C adatom can go over a shallow barrier to a more stable adsorption position, namely Mo-1 site with $E_b$=3.28 eV at the center of three Mo atoms below the plane of S atoms. In \ref{fig:C} we present a detailed analysis of these two adsorption sites. For Mo-1 site, $sp^2$ hybridized orbitals of C adatom and $d_{xy}$ and $d_{x^{2}-y^{2}}$ orbitals of surrounding 3 Mo atoms form three bonds, but the system remains nonmagnetic. In the case of S-3 site, $sp^2$ hybridized orbitals of C adatom form three bonds with the orbitals of surrounding three S atoms of 1H-MoS$_2$. The interaction of $p_z$ orbital of C adatom with the underlying Mo $d_{z^2}$ orbitals results in a significant charge transfer from Mo to C adatom. This creates  a local magnetic moment of $\mu$=2 $\mu_{B}$ on Mo atom. The charge density contour plots in \ref{fig:C} (b) and (c) in a perpendicular and lateral planes show the bonding configuration. Recently a peculiar growth process, where carbon adatoms adsorbed to graphene readily diffuse above room temperature and nucleate segments of linear carbon chains (CACs) attached on graphene is predicted.\cite{graph+C} The diffusion barrier they calculate on graphene is reported as 0.37 eV. The energy difference of C adatom between different adsorption sites, namely S-3 and S-4, is 0.45 eV and comparable to that of C atom on graphene. Here we went beyond the adsorption of individual atoms and addressed the question whether carbon adatoms can also lead to the formation of CACs on 1H-MoS$_2$. Different from the single C adatom adsorption on 1H-MoS$_2$, CACs (starting from C$_2$) prefer to adsorb on the top of S atoms as shown in \ref{fig:C} (d) and (e). When attached to 1H-MoS$_2$, CACs favor cumulene like structures with slightly alternating bonds and with a different kind of even-odd disparity. Therefore the top-site geometry occurs due to the double $sp$+$\pi$-bonding within cumulene. Similar to the case in graphene, the binding energy of CACs having even number of C atoms are greater than those having odd number of C atoms. Even though CAC on graphene are nonmagnetic structures, odd number CACs on 1H-MoS$_2$ have a magnetic moment of 2 $\mu_B$ arising from C atoms at the edges of CAC.

Since the atomic radii of other group 4A elements, Si and Ge, are larger than that of C adatom, their adsorption geometry differs from the adsorption site of C at the S-3 site. Si and Ge can only be adsorbed on the S-plane at the S-4 site and attain a local magnetic moment of $\mu=$ 2 $\mu_B$. This is a crucial result, since magnetic properties to MoS$_2$ monolayer are implemented through nonmagnetic adatoms.

\begin{center}
\begin{figure}
\includegraphics[width=8.25cm]{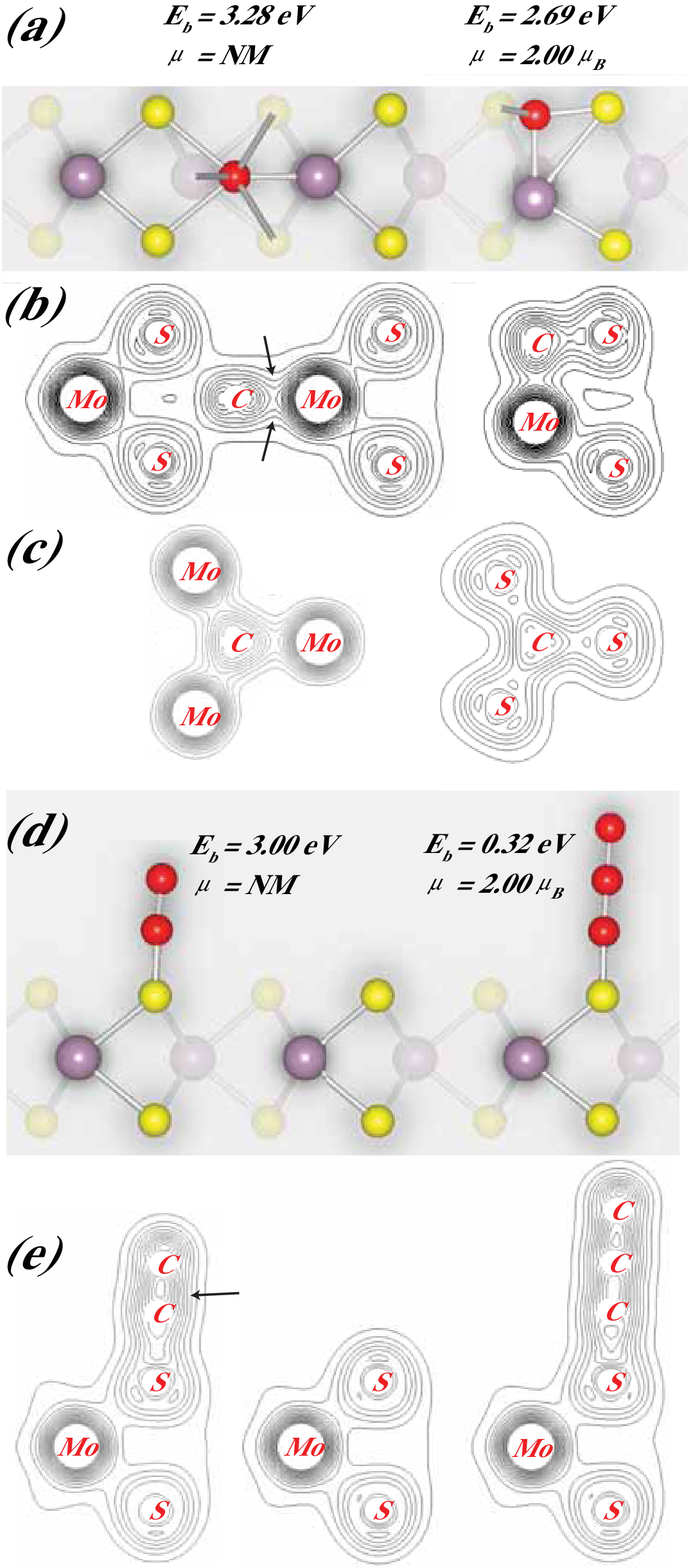}
\caption{Analysis of bonding configuration of C adatoms on 1H-MoS$_2$. (a) Geometries of single C adatom adsorbed in the Mo-plane (left) and in the S-plane (right). Adatoms, host Mo and S atoms are represented by red (medium-dark), purple (large-gray) and yellow (small-light) balls, respectively. (b) Contour plots of total charge density of plane passing through atoms and bonds highlighted (not shaded) in (a). (c) Contour plots of total charge density on the horizontal plane passing through Mo-C and S-C bonds parallel to 1H-MoS$_2$. (d) Adsorption geometry and energetics of C$_2$ and C$_3$ on 1H-MoS$_2$. (e) Contour plots of total charge density on the vertical plane passing through atoms and bonds emphasized (not shaded) in (d). The direction of the arrows indicate the increasing charge density.}
\label{fig:C}
\end{figure}
\par\end{center}

The excess charge on the adatom (which is denoted as $\rho^*$ in \ref{tab:2da}), as well as the position of the highest localized state in the gap are correlated with its electronegativity. For example, among C, Si and Ge, C has the highest electronegativity. While C has negative excess charge, Si and Ge becomes positively charged when adsorbed. Calculated dipole moment on the system of C adatom adsorbed to S-layer is diminished due to the local reconstruction. While adsorbed C, O and S adatoms have excess electrons, all other adatoms in \ref{tab:2da} are positively charged. Vanadium and W adsorbed at S-2 site have significant positive charge, but both induce minute electric dipole moment due to the local reconstruction. Finally, it should be noted that excess charging of surfaces of 1H-MoS$_2$ by a higher level of coverage of specific adatoms can improve its functionalities in tribology.

\section{Vacancy Defects}

It is known that the vacancy defect in 2D graphene,\cite{esquinazi,Iijima,yazyev,guinea,brey2} graphene nanoribbons,\cite{brey1,delik} 2D graphane\cite{sahingraphane} and graphane nanoribbons\cite{sahinprb} give rise to crucial changes in the electronic and magnetic structure. According to Lieb's theorem,\cite{lieb} the net magnetic moment of the structure occurs due to the difference in the number of atoms belonging to different sublattices A and B, namely $\mu= (N_{A}-N_{B})\mu_{B}$. While magnetic moments calculated using DFT for vacancies in 2D graphene and armchair graphene nanoribbons generally confirmed Lieb's theorem, results are diversified for vacancies in zigzag graphene nanoribbons\cite{brey1,delik} due to the coupling with magnetic edge states. In the case of 1H-MoS$_{2}$, Lieb's theorem cannot be directly applicable, even if alternating Mo and S units form a honeycomb structure. We investigated five different types of vacancy defects, namely Mo-, S-vacancy, MoS-, S$_2$-divacancy and MoS$_2$-triple vacancy, which are formed in 2D 1H-MoS$_2$. All structures are optimized upon the creation of a particular type of vacancy. Vacancy energies, $E_V$, are calculated by subtracting the total energy of the perfect structure (without vacancy) from the sum of the total energy of a structure having a particular type of vacancy and the total energy(ies) of missing atoms in the vacancy defect. Here all structures are optimized in their ground states (whether magnetic or nonmagnetic). Positive $E_V$ indicates that the formation of vacancy defect is an endothermic process. In \ref{tab:vacancy} vacancy energies as defined above and their magnetic ground states are presented. We note that the equilibrium concentrations of vacancies are usually very low owing their high formation energies. Nonetheless, new techniques have been developed to create defects, which makes also the generation of nanomeshes vacancies  possible.\cite{bai,lahiri,balog}

\begin{table}
\caption{Calculated vacancy energies $E_V$ (in eV), magnetic moments $\mu$ (in $\mu_B$) of five different types of vacancy defects, Mo, MoS, MoS$_2$, S, S$_2$ in (7x7) supercell of 1H-MoS$_2$. NM stands for nonmagnetic state with net $\mu$=0 $\mu_{B}$. $E_{i}$s denote the energies of localized states in the band gap measured from the top of the valence bands (in eV). The occupied ones are indicated by bold numerals and their spin alignments are denoted by either $\uparrow$ or $\downarrow$. States without the indication of spin alignment are nonmagnetic.} \label{tab:vacancy}
\begin{center}
\begin{tabular}{ccccccc}
\hline  \hline
& Mo & MoS& MoS$_{2}$ & S & S$_{2}$\\
& $E_V$-$\mu$ & $E_V$-$\mu$ & $E_V$-$\mu$  & $E_V$-$\mu$ & $E_V$-$\mu$ \\
\hline
2D  &  13.44-NM & 17.36-NM & 22.63-2.00 & 5.89-NM & 11.74-NM  & \\
\hline
E$_1$ & \textbf{0.26} & \textbf{0.15} & \textbf{0.15} $\uparrow$ &\textbf{ 0.12} & \textbf{0.12} \tabularnewline
E$_2$ & \textbf{0.27} & \textbf{0.25} & \textbf{0.16} $\uparrow$ & 1.22 & 1.17 \tabularnewline
E$_3$ & 0.61  & 0.54 & \textbf{0.25} $\downarrow$ & 1.23 & 1.19 \tabularnewline
E$_4$ & 1.02 & 0.69 & \textbf{0.28} $\uparrow$ & & \tabularnewline
E$_5$ &      & 1.21 & 0.53 $\downarrow$ & & \tabularnewline
E$_6$ &      & 1.36 & 0.73 $\uparrow$ & & \tabularnewline
\hline \hline
\end{tabular}
\end{center}
\end{table}

We carried out calculations on vacancy defects of Mo, S and S$_2$, in the (2x2), (4x4) and (7x7) supercells of 1H-MoS$_2$ structure. Even if relatively larger sizes of supercells are not affected, the size of supercell is contracted upon creating a vacancy defect in the (2x2) supercell. In addition to these, we also consider MoS-divacancy and MoS$_2$-triple vacancies in (7x7) supercells. For all these supercell sizes including the (7x7) supercell, we have carried out calculations with many different initial magnetic moment configurations, but all of the vacancy defects, except MoS$_2$ triple vacancy, the magnetic moment on any of the atoms in the supercell vanished. To justify these results, we also repeated the calculations using LDA+PAW, which also results in a magnetic state for MoS$_2$-triple vacancy having a net magnetic moment of $\mu_{T} = 2 \mu_B$. We seek the origin of magnetic states in the charge transfer between Mo and S and hence examined charge densities around the vacancy defects using Bader analysis.\cite{Bader} In case of S and S$_2$ vacancy defects, the excess (positive) charge on the nearest Mo adatoms around the vacancy slightly decreases, and hence does not cause any magnetic moment. The charge transfers and resulting excess charges around Mo and MoS$_2$ vacancies are affected even up to third nearest neighbor atoms. For example S atoms around Mo vacancy have 0.1 electrons less charge, since the third Mo atom, which provides excess electrons to S is missing. However these S atoms surrounding Mo vacancy receive more charge from nearest two Mo atoms. Similar cases also investigated for MoS vacancy defects. Nonetheless, the charge transfers at the close proximity of S, S$_2$, Mo and MoS vacancies are not significant as compared with those of the perfect structure and consequently do not lead to a magnetic state. Interestingly, the disturbances in the charge transfer due to MoS$_2$-triple vacancy are significant and result in the magnetic state at the close proximity of the defect.

\begin{center}
\begin{figure}
\includegraphics[width=8.25cm]{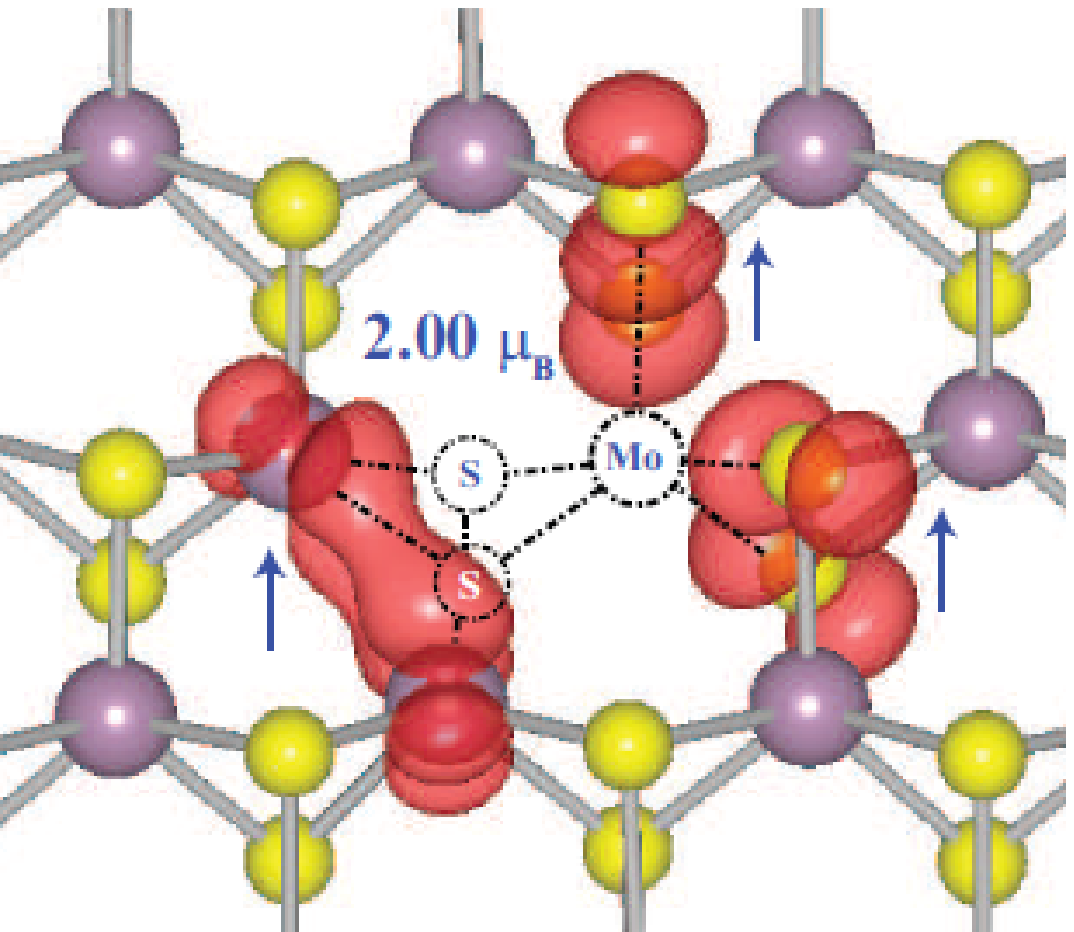}
\caption{Isosurfaces of difference charge density of MoS$_2$ vacancy defect in the (7x7) supercell of 1H-MoS$_2$. Dashed atoms and bonds are vacant sites. Difference charge density is obtained from the difference of spin-up and spin-down charge densities. ($\Delta\rho_{\uparrow,\downarrow} = \rho_{\uparrow}-\rho_{\downarrow}$) The total magnetic moment is calculated as 2 $\mu_{B}$. Up arrow indicate the excess spin-up charge. Isosurface value is taken as 3x10$^{-3}$ eV/\AA$^{3}$. }
\label{fig:vacancy}
\end{figure}
\end{center}

In \ref{fig:vacancy}, we present the isosurfaces of the difference of charge density of spin-up and spin-down states (i.e. $\Delta \rho_{\uparrow,\downarrow} = \rho_{\uparrow}-\rho_{\downarrow}$ at the close proximity of MoS$_2$ vacancy. After reconstruction around the vacancy, 2 Mo and 4 S atoms have dangling bonds. In this case, Mo atoms having dangling bonds are also less positively charged and S atoms having dangling bonds are less negatively charged as compared with those of perfect MoS$_2$. However, in comparison with S, S$_2$, Mo and MoS vacancies the amount of charge transfers here are almost doubled to cause to significant disturbances and spin polarization. The total magnetic moment of 2 $\mu_{B}$ are originated equally from $d_{yz}$ and $d_{zx}$ orbitals of Mo and $p$ orbitals of S which have dangling bonds as seen in \ref{fig:vacancy}. The nonmagnetic state is $\sim 130$ meV energetically less favorable. These results are also consisted with the vacancy defects in armchair edged MoS$_2$ nanoribbons.\cite{ataca} Electronically, vacancy defects give rise to states in the band gap, which are localized at atoms around the vacancy (see \ref{tab:vacancy}). The band gap, as well as the electronic properties of 1H-MoS$_2$ are modified by these states.

\section{Discussion and Conclusions}

The adsorption of adatoms and creation of vacancy defects in 2D single layer of MoS$_2$ honeycomb structure have crucial effects on the electronic and magnetic properties. We found that several adatoms can be adsorbed readily at diverse sites with significant binding energy. In this respect, MoS$_2$ appears to be a material, which is suitable for functionalization. While two dimensional, single layer MoS$_2$ is a direct band gap semiconductor, magnetic or nonmagnetic localized gap states due to adatoms occur in the band gap and expand the capacity of this material in nanoelectronics and promise future applications. Additionally, two dimensional sheets of MoS$_2$ can attain a local magnetic moments through adsorption of $3d$-transition metal atoms, as well as Si and Ge. Carbon being of particular interest, we also examined the adsorption of C$_2$ and linear C$_3$.
Significant amount of charge is transferred to (or from) adatom. Excess surface charge at higher coverage of adatoms can improve tribological and catalytic properties of 1H-MoS$_2$. While vacancy defects of S, S$_2$, Mo and MoS created in 2D 1H-MoS$_2$ do not induce any magnetic moment, the creation of MoS$_2$ triple vacancy results in a significant magnetic moment in the system. Briefly, functionalization of MoS$_2$ honeycomb structure through adatom adsorption and vacancy creation appears to be a promising way to extend the applications of MoS$_2$. Since the interlayer interaction in graphitic MoS$_2$ is weak van der Waals, present results are also relevant for the surfaces of 3D MoS$_2$.

\begin{acknowledgement}

This work is supported by TUBITAK through Grant No:104T537 and Grant No: 108T234. Part of the computational resources have been provided by UYBHM at Istanbul Technical University. S.C. acknowledges TUBA for partial support. We thank the DEISA Consortium (www.deisa.eu), funded through the EU FP7 project RI-222919, for support within the DEISA Extreme Computing Initiative.

\end{acknowledgement}

\bibliography{functionalization-acs}

\providecommand*\mcitethebibliography{\thebibliography}
\csname @ifundefined\endcsname{endmcitethebibliography}
  {\let\endmcitethebibliography\endthebibliography}{}
\begin{mcitethebibliography}{78}
\providecommand*\natexlab[1]{#1}
\providecommand*\mciteSetBstSublistMode[1]{}
\providecommand*\mciteSetBstMaxWidthForm[2]{}
\providecommand*\mciteBstWouldAddEndPuncttrue
  {\def\EndOfBibitem{\unskip.}}
\providecommand*\mciteBstWouldAddEndPunctfalse
  {\let\EndOfBibitem\relax}
\providecommand*\mciteSetBstMidEndSepPunct[3]{}
\providecommand*\mciteSetBstSublistLabelBeginEnd[3]{}
\providecommand*\EndOfBibitem{}
\mciteSetBstSublistMode{f}
\mciteSetBstMaxWidthForm{subitem}{(\alph{mcitesubitemcount})}
\mciteSetBstSublistLabelBeginEnd
  {\mcitemaxwidthsubitemform\space}
  {\relax}
  {\relax}

\bibitem[JOENSEN et~al.({1986})JOENSEN, FRINDT, and MORRISON]{Joenson}
JOENSEN,~P.; FRINDT,~R.; MORRISON,~S. \emph{{MATERIALS RESEARCH BULLETIN}}
  \textbf{{1986}}, \emph{{21}}, {457--461}\relax
\mciteBstWouldAddEndPuncttrue
\mciteSetBstMidEndSepPunct{\mcitedefaultmidpunct}
{\mcitedefaultendpunct}{\mcitedefaultseppunct}\relax
\EndOfBibitem
\bibitem[Helveg et~al.({2000})Helveg, Lauritsen, Laegsgaard, Stensgaard,
  Norskov, Clausen, Topsoe, and Besenbacher]{Helveg}
Helveg,~S.; Lauritsen,~J.; Laegsgaard,~E.; Stensgaard,~I.; Norskov,~J.;
  Clausen,~B.; Topsoe,~H.; Besenbacher,~F. \emph{{PHYSICAL REVIEW LETTERS}}
  \textbf{{2000}}, \emph{{84}}, {951--954}\relax
\mciteBstWouldAddEndPuncttrue
\mciteSetBstMidEndSepPunct{\mcitedefaultmidpunct}
{\mcitedefaultendpunct}{\mcitedefaultseppunct}\relax
\EndOfBibitem
\bibitem[Lee et~al.({2010})Lee, Yan, Brus, Heinz, Hone, and Ryu]{Lee}
Lee,~C.; Yan,~H.; Brus,~L.~E.; Heinz,~T.~F.; Hone,~J.; Ryu,~S. \emph{{ACS
  NANO}} \textbf{{2010}}, \emph{{4}}, {2695--2700}\relax
\mciteBstWouldAddEndPuncttrue
\mciteSetBstMidEndSepPunct{\mcitedefaultmidpunct}
{\mcitedefaultendpunct}{\mcitedefaultseppunct}\relax
\EndOfBibitem
\bibitem[Novoselov et~al.({2005})Novoselov, Jiang, Schedin, Booth, Khotkevich,
  Morozov, and Geim]{novo2}
Novoselov,~K.; Jiang,~D.; Schedin,~F.; Booth,~T.; Khotkevich,~V.; Morozov,~S.;
  Geim,~A. \emph{{PROCEEDINGS OF THE NATIONAL ACADEMY OF SCIENCES OF THE UNITED
  STATES OF AMERICA}} \textbf{{2005}}, \emph{{102}}, {10451--10453}\relax
\mciteBstWouldAddEndPuncttrue
\mciteSetBstMidEndSepPunct{\mcitedefaultmidpunct}
{\mcitedefaultendpunct}{\mcitedefaultseppunct}\relax
\EndOfBibitem
\bibitem[Mak et~al.(2010)Mak, Lee, Hone, Shan, and Heinz]{Mak}
Mak,~K.~F.; Lee,~C.; Hone,~J.; Shan,~J.; Heinz,~T.~F. \emph{Phys. Rev. Lett.}
  \textbf{2010}, \emph{105}, 136805\relax
\mciteBstWouldAddEndPuncttrue
\mciteSetBstMidEndSepPunct{\mcitedefaultmidpunct}
{\mcitedefaultendpunct}{\mcitedefaultseppunct}\relax
\EndOfBibitem
\bibitem[Coleman et~al.(2011)Coleman, Lotya, O'Neill, Bergin, King, Khan,
  Young, Gaucher, De, Smith, Shvets, Arora, Stanton, Kim, Lee, Kim, Duesberg,
  Hallam, Boland, Wang, Donegan, Grunlan, Moriarty, Shmeliov, Nicholls,
  Perkins, Grieveson, Theuwissen, McComb, Nellist, and Nicolosi]{Coleman}
Coleman,~J.~N. et~al.  \emph{Science} \textbf{2011}, \emph{331}, 568--571\relax
\mciteBstWouldAddEndPuncttrue
\mciteSetBstMidEndSepPunct{\mcitedefaultmidpunct}
{\mcitedefaultendpunct}{\mcitedefaultseppunct}\relax
\EndOfBibitem
\bibitem[Li et~al.({2008})Li, Zhou, Zhang, and Chen]{Li}
Li,~Y.; Zhou,~Z.; Zhang,~S.; Chen,~Z. \emph{{JOURNAL OF THE AMERICAN CHEMICAL
  SOCIETY}} \textbf{{2008}}, \emph{{130}}, {16739--16744}\relax
\mciteBstWouldAddEndPuncttrue
\mciteSetBstMidEndSepPunct{\mcitedefaultmidpunct}
{\mcitedefaultendpunct}{\mcitedefaultseppunct}\relax
\EndOfBibitem
\bibitem[Lebegue and Eriksson({2009})Lebegue, and Eriksson]{Lebegue}
Lebegue,~S.; Eriksson,~O. \emph{{PHYSICAL REVIEW B}} \textbf{{2009}},
  \emph{{79}}, {115409}\relax
\mciteBstWouldAddEndPuncttrue
\mciteSetBstMidEndSepPunct{\mcitedefaultmidpunct}
{\mcitedefaultendpunct}{\mcitedefaultseppunct}\relax
\EndOfBibitem
\bibitem[Botello-Mendez et~al.({2009})Botello-Mendez, Lopez-Urias, Terrones,
  and Terrones]{Mendez}
Botello-Mendez,~A.~R.; Lopez-Urias,~F.; Terrones,~M.; Terrones,~H.
  \emph{{NANOTECHNOLOGY}} \textbf{{2009}}, \emph{{20}}, {325703}\relax
\mciteBstWouldAddEndPuncttrue
\mciteSetBstMidEndSepPunct{\mcitedefaultmidpunct}
{\mcitedefaultendpunct}{\mcitedefaultseppunct}\relax
\EndOfBibitem
\bibitem[Bollinger et~al.({2001})Bollinger, Lauritsen, Jacobsen, Norskov,
  Helveg, and Besenbacher]{Bollinger}
Bollinger,~M.; Lauritsen,~J.; Jacobsen,~K.; Norskov,~J.; Helveg,~S.;
  Besenbacher,~F. \emph{{PHYSICAL REVIEW LETTERS}} \textbf{{2001}},
  \emph{{87}}, {196803}\relax
\mciteBstWouldAddEndPuncttrue
\mciteSetBstMidEndSepPunct{\mcitedefaultmidpunct}
{\mcitedefaultendpunct}{\mcitedefaultseppunct}\relax
\EndOfBibitem
\bibitem[Ataca et~al.(2011)Ataca, Sahin, Akturk, and Ciraci]{ataca}
Ataca,~C.; Sahin,~H.; Akturk,~E.; Ciraci,~S. \emph{The Journal of Physical
  Chemistry C} \textbf{2011}, \emph{115}, 3934--3941\relax
\mciteBstWouldAddEndPuncttrue
\mciteSetBstMidEndSepPunct{\mcitedefaultmidpunct}
{\mcitedefaultendpunct}{\mcitedefaultseppunct}\relax
\EndOfBibitem
\bibitem[Li and Galli(2007)Li, and Galli]{Galli}
Li,~T.; Galli,~G. \emph{The Journal of Physical Chemistry C} \textbf{2007},
  \emph{111}, 16192--16196\relax
\mciteBstWouldAddEndPuncttrue
\mciteSetBstMidEndSepPunct{\mcitedefaultmidpunct}
{\mcitedefaultendpunct}{\mcitedefaultseppunct}\relax
\EndOfBibitem
\bibitem[MARTIN et~al.({1993})MARTIN, DONNET, LEMOGNE, and EPICIER]{Martin}
MARTIN,~J.; DONNET,~C.; LEMOGNE,~T.; EPICIER,~T. \emph{{PHYSICAL REVIEW B}}
  \textbf{{1993}}, \emph{{48}}, {10583--10586}\relax
\mciteBstWouldAddEndPuncttrue
\mciteSetBstMidEndSepPunct{\mcitedefaultmidpunct}
{\mcitedefaultendpunct}{\mcitedefaultseppunct}\relax
\EndOfBibitem
\bibitem[Hoelscher et~al.({2008})Hoelscher, Ebeling, and Schwarz]{Holscher}
Hoelscher,~H.; Ebeling,~D.; Schwarz,~U.~D. \emph{{PHYSICAL REVIEW LETTERS}}
  \textbf{{2008}}, \emph{{101}}, {246105}\relax
\mciteBstWouldAddEndPuncttrue
\mciteSetBstMidEndSepPunct{\mcitedefaultmidpunct}
{\mcitedefaultendpunct}{\mcitedefaultseppunct}\relax
\EndOfBibitem
\bibitem[Lee et~al.({2010})Lee, Li, Kalb, Liu, Berger, Carpick, and Hone]{Lee2}
Lee,~C.; Li,~Q.; Kalb,~W.; Liu,~X.-Z.; Berger,~H.; Carpick,~R.~W.; Hone,~J.
  \emph{{SCIENCE}} \textbf{{2010}}, \emph{{328}}, {76--80}\relax
\mciteBstWouldAddEndPuncttrue
\mciteSetBstMidEndSepPunct{\mcitedefaultmidpunct}
{\mcitedefaultendpunct}{\mcitedefaultseppunct}\relax
\EndOfBibitem
\bibitem[Liang et~al.({2008})Liang, Sawyer, Perry, Sinnott, and
  Phillpot]{Liang}
Liang,~T.; Sawyer,~W.~G.; Perry,~S.~S.; Sinnott,~S.~B.; Phillpot,~S.~R.
  \emph{{PHYSICAL REVIEW B}} \textbf{{2008}}, \emph{{77}}, {104105}\relax
\mciteBstWouldAddEndPuncttrue
\mciteSetBstMidEndSepPunct{\mcitedefaultmidpunct}
{\mcitedefaultendpunct}{\mcitedefaultseppunct}\relax
\EndOfBibitem
\bibitem[Onodera et~al.({2009})Onodera, Morita, Suzuki, Koyama, Tsuboi,
  Hatakeyama, Endou, Takaba, Kubo, Dassenoy, Minfray, Joly-Pottuz, Martin, and
  Miyamoto]{Onodra}
Onodera,~T.; Morita,~Y.; Suzuki,~A.; Koyama,~M.; Tsuboi,~H.; Hatakeyama,~N.;
  Endou,~A.; Takaba,~H.; Kubo,~M.; Dassenoy,~F.; Minfray,~C.; Joly-Pottuz,~L.;
  Martin,~J.-M.; Miyamoto,~A. \emph{{JOURNAL OF PHYSICAL CHEMISTRY B}}
  \textbf{{2009}}, \emph{{113}}, {16526--16536}\relax
\mciteBstWouldAddEndPuncttrue
\mciteSetBstMidEndSepPunct{\mcitedefaultmidpunct}
{\mcitedefaultendpunct}{\mcitedefaultseppunct}\relax
\EndOfBibitem
\bibitem[Hinnemann et~al.({2005})Hinnemann, Moses, Bonde, Jorgensen, Nielsen,
  Horch, Chorkendorff, and Norskov]{Hinnemann}
Hinnemann,~B.; Moses,~P.; Bonde,~J.; Jorgensen,~K.; Nielsen,~J.; Horch,~S.;
  Chorkendorff,~I.; Norskov,~J. \emph{{JOURNAL OF THE AMERICAN CHEMICAL
  SOCIETY}} \textbf{{2005}}, \emph{{127}}, {5308--5309}\relax
\mciteBstWouldAddEndPuncttrue
\mciteSetBstMidEndSepPunct{\mcitedefaultmidpunct}
{\mcitedefaultendpunct}{\mcitedefaultseppunct}\relax
\EndOfBibitem
\bibitem[Jaramillo et~al.({2007})Jaramillo, Jorgensen, Bonde, Nielsen, Horch,
  and Chorkendorff]{Jaramillo}
Jaramillo,~T.~F.; Jorgensen,~K.~P.; Bonde,~J.; Nielsen,~J.~H.; Horch,~S.;
  Chorkendorff,~I. \emph{{SCIENCE}} \textbf{{2007}}, \emph{{317}},
  {100--102}\relax
\mciteBstWouldAddEndPuncttrue
\mciteSetBstMidEndSepPunct{\mcitedefaultmidpunct}
{\mcitedefaultendpunct}{\mcitedefaultseppunct}\relax
\EndOfBibitem
\bibitem[Lauritsen et~al.({2001})Lauritsen, Helveg, Laegsgaard, Stensgaard,
  Clausen, Topsoe, and Besenbacher]{HDS2}
Lauritsen,~J.; Helveg,~S.; Laegsgaard,~E.; Stensgaard,~I.; Clausen,~B.;
  Topsoe,~H.; Besenbacher,~E. \emph{{JOURNAL OF CATALYSIS}} \textbf{{2001}},
  \emph{{197}}, {1--5}\relax
\mciteBstWouldAddEndPuncttrue
\mciteSetBstMidEndSepPunct{\mcitedefaultmidpunct}
{\mcitedefaultendpunct}{\mcitedefaultseppunct}\relax
\EndOfBibitem
\bibitem[Lauritsen et~al.({2004})Lauritsen, Bollinger, Laegsgaard, Jacobsen,
  Norskov, Clausen, Topsoe, and Besenbacher]{HDS3}
Lauritsen,~J.; Bollinger,~M.; Laegsgaard,~E.; Jacobsen,~K.; Norskov,~J.;
  Clausen,~B.; Topsoe,~H.; Besenbacher,~F. \emph{{JOURNAL OF CATALYSIS}}
  \textbf{{2004}}, \emph{{221}}, {510--522}\relax
\mciteBstWouldAddEndPuncttrue
\mciteSetBstMidEndSepPunct{\mcitedefaultmidpunct}
{\mcitedefaultendpunct}{\mcitedefaultseppunct}\relax
\EndOfBibitem
\bibitem[Lauritsen et~al.({2007})Lauritsen, Kibsgaard, Olesen, Moses,
  Hinnemann, Helveg, Norskov, Clausen, Topsoe, Laegsgaard, and
  Besenbacher]{HDS4}
Lauritsen,~J.~V.; Kibsgaard,~J.; Olesen,~G.~H.; Moses,~P.~G.; Hinnemann,~B.;
  Helveg,~S.; Norskov,~J.~K.; Clausen,~B.~S.; Topsoe,~H.; Laegsgaard,~E.;
  Besenbacher,~F. \emph{{JOURNAL OF CATALYSIS}} \textbf{{2007}}, \emph{{249}},
  {220--233}\relax
\mciteBstWouldAddEndPuncttrue
\mciteSetBstMidEndSepPunct{\mcitedefaultmidpunct}
{\mcitedefaultendpunct}{\mcitedefaultseppunct}\relax
\EndOfBibitem
\bibitem[Moses et~al.({2007})Moses, Hinnemann, Topsoe, and Norskov]{HDS5}
Moses,~P.~G.; Hinnemann,~B.; Topsoe,~H.; Norskov,~J.~K. \emph{{JOURNAL OF
  CATALYSIS}} \textbf{{2007}}, \emph{{248}}, {188--203}\relax
\mciteBstWouldAddEndPuncttrue
\mciteSetBstMidEndSepPunct{\mcitedefaultmidpunct}
{\mcitedefaultendpunct}{\mcitedefaultseppunct}\relax
\EndOfBibitem
\bibitem[Raybaud et~al.({2000})Raybaud, Hafner, Kresse, Kasztelan, and
  Toulhoat]{HDS6}
Raybaud,~P.; Hafner,~J.; Kresse,~G.; Kasztelan,~S.; Toulhoat,~H. \emph{{JOURNAL
  OF CATALYSIS}} \textbf{{2000}}, \emph{{189}}, {129--146}\relax
\mciteBstWouldAddEndPuncttrue
\mciteSetBstMidEndSepPunct{\mcitedefaultmidpunct}
{\mcitedefaultendpunct}{\mcitedefaultseppunct}\relax
\EndOfBibitem
\bibitem[Sun et~al.({2004})Sun, Nelson, and Adjaye]{HDS7}
Sun,~M.; Nelson,~A.; Adjaye,~J. \emph{{JOURNAL OF CATALYSIS}} \textbf{{2004}},
  \emph{{226}}, {32--40}\relax
\mciteBstWouldAddEndPuncttrue
\mciteSetBstMidEndSepPunct{\mcitedefaultmidpunct}
{\mcitedefaultendpunct}{\mcitedefaultseppunct}\relax
\EndOfBibitem
\bibitem[Todorova et~al.({2007})Todorova, Prins, and Weber]{HDS8}
Todorova,~T.; Prins,~R.; Weber,~T. \emph{{JOURNAL OF CATALYSIS}}
  \textbf{{2007}}, \emph{{246}}, {109--117}\relax
\mciteBstWouldAddEndPuncttrue
\mciteSetBstMidEndSepPunct{\mcitedefaultmidpunct}
{\mcitedefaultendpunct}{\mcitedefaultseppunct}\relax
\EndOfBibitem
\bibitem[KLINE et~al.({1982})KLINE, KAM, ZIEGLER, and PARKINSON]{Kline}
KLINE,~G.; KAM,~K.; ZIEGLER,~R.; PARKINSON,~B. \emph{{SOLAR ENERGY MATERIALS}}
  \textbf{{1982}}, \emph{{6}}, {337--350}\relax
\mciteBstWouldAddEndPuncttrue
\mciteSetBstMidEndSepPunct{\mcitedefaultmidpunct}
{\mcitedefaultendpunct}{\mcitedefaultseppunct}\relax
\EndOfBibitem
\bibitem[Wilcoxon et~al.({1999})Wilcoxon, Thurston, and Martin]{Wilcoson}
Wilcoxon,~J.; Thurston,~T.; Martin,~J. \emph{{NANOSTRUCTURED MATERIALS}}
  \textbf{{1999}}, \emph{{12}}, {993--997}, {4th International Conference on
  Nanostructured Materials (NANO 98), STOCKHOLM, SWEDEN, JUN 14-19, 1998}\relax
\mciteBstWouldAddEndPuncttrue
\mciteSetBstMidEndSepPunct{\mcitedefaultmidpunct}
{\mcitedefaultendpunct}{\mcitedefaultseppunct}\relax
\EndOfBibitem
\bibitem[Splendiani et~al.(2010)Splendiani, Sun, Zhang, Li, Kim, Chim, Galli,
  and Wang]{splendiani}
Splendiani,~A.; Sun,~L.; Zhang,~Y.; Li,~T.; Kim,~J.; Chim,~C.-Y.; Galli,~G.;
  Wang,~F. \emph{Nano Letters} \textbf{2010}, \emph{10}, 1271--1275, PMID:
  20229981\relax
\mciteBstWouldAddEndPuncttrue
\mciteSetBstMidEndSepPunct{\mcitedefaultmidpunct}
{\mcitedefaultendpunct}{\mcitedefaultseppunct}\relax
\EndOfBibitem
\bibitem[Radisavljevic et~al.({2011})Radisavljevic, Radenovic, Brivio,
  Giacometti, and Kis]{Radi}
Radisavljevic,~B.; Radenovic,~A.; Brivio,~J.; Giacometti,~V.; Kis,~A.
  \emph{{NATURE NANOTECHNOLOGY}} \textbf{{2011}}, \emph{{6}}, {147--150}\relax
\mciteBstWouldAddEndPuncttrue
\mciteSetBstMidEndSepPunct{\mcitedefaultmidpunct}
{\mcitedefaultendpunct}{\mcitedefaultseppunct}\relax
\EndOfBibitem
\bibitem[Raybaud et~al.({1998})Raybaud, Hafner, Kresse, and Toulhoat]{Raybaud}
Raybaud,~P.; Hafner,~J.; Kresse,~G.; Toulhoat,~H. \emph{{PHYSICAL REVIEW
  LETTERS}} \textbf{{1998}}, \emph{{80}}, {1481--1484}\relax
\mciteBstWouldAddEndPuncttrue
\mciteSetBstMidEndSepPunct{\mcitedefaultmidpunct}
{\mcitedefaultendpunct}{\mcitedefaultseppunct}\relax
\EndOfBibitem
\bibitem[Moses et~al.({2009})Moses, Mortensen, Lundqvist, and Norskov]{Moses}
Moses,~P.~G.; Mortensen,~J.~J.; Lundqvist,~B.~I.; Norskov,~J.~K. \emph{{JOURNAL
  OF CHEMICAL PHYSICS}} \textbf{{2009}}, \emph{{130}}, {104709}\relax
\mciteBstWouldAddEndPuncttrue
\mciteSetBstMidEndSepPunct{\mcitedefaultmidpunct}
{\mcitedefaultendpunct}{\mcitedefaultseppunct}\relax
\EndOfBibitem
\bibitem[Huang and Cho({2009})Huang, and Cho]{Huang}
Huang,~M.; Cho,~K. \emph{{JOURNAL OF PHYSICAL CHEMISTRY C}} \textbf{{2009}},
  \emph{{113}}, {5238--5243}\relax
\mciteBstWouldAddEndPuncttrue
\mciteSetBstMidEndSepPunct{\mcitedefaultmidpunct}
{\mcitedefaultendpunct}{\mcitedefaultseppunct}\relax
\EndOfBibitem
\bibitem[Fuhr et~al.({2004})Fuhr, Saul, and Sofo]{Fuhr}
Fuhr,~J.; Saul,~A.; Sofo,~J. \emph{{PHYSICAL REVIEW LETTERS}} \textbf{{2004}},
  \emph{{92}}, {026802}\relax
\mciteBstWouldAddEndPuncttrue
\mciteSetBstMidEndSepPunct{\mcitedefaultmidpunct}
{\mcitedefaultendpunct}{\mcitedefaultseppunct}\relax
\EndOfBibitem
\bibitem[Ivanovskaya et~al.({2008})Ivanovskaya, Zobelli, Gloter, Brun, Serin,
  and Colliex]{Ivanovskaya}
Ivanovskaya,~V.~V.; Zobelli,~A.; Gloter,~A.; Brun,~N.; Serin,~V.; Colliex,~C.
  \emph{{PHYSICAL REVIEW B}} \textbf{{2008}}, \emph{{78}}, {134104}\relax
\mciteBstWouldAddEndPuncttrue
\mciteSetBstMidEndSepPunct{\mcitedefaultmidpunct}
{\mcitedefaultendpunct}{\mcitedefaultseppunct}\relax
\EndOfBibitem
\bibitem[He et~al.({2010})He, Wu, Sa, Li, and Wei]{He}
He,~J.; Wu,~K.; Sa,~R.; Li,~Q.; Wei,~Y. \emph{{APPLIED PHYSICS LETTERS}}
  \textbf{{2010}}, \emph{{96}}, {082504}\relax
\mciteBstWouldAddEndPuncttrue
\mciteSetBstMidEndSepPunct{\mcitedefaultmidpunct}
{\mcitedefaultendpunct}{\mcitedefaultseppunct}\relax
\EndOfBibitem
\bibitem[BLOCHL({1994})]{paw}
BLOCHL,~P. \emph{{PHYSICAL REVIEW B}} \textbf{{1994}}, \emph{{50}},
  {17953--17979}\relax
\mciteBstWouldAddEndPuncttrue
\mciteSetBstMidEndSepPunct{\mcitedefaultmidpunct}
{\mcitedefaultendpunct}{\mcitedefaultseppunct}\relax
\EndOfBibitem
\bibitem[PERDEW et~al.({1992})PERDEW, CHEVARY, VOSKO, JACKSON, PEDERSON, SINGH,
  and FIOLHAIS]{pw91}
PERDEW,~J.; CHEVARY,~J.; VOSKO,~S.; JACKSON,~K.; PEDERSON,~M.; SINGH,~D.;
  FIOLHAIS,~C. \emph{{PHYSICAL REVIEW B}} \textbf{{1992}}, \emph{{46}},
  {6671--6687}\relax
\mciteBstWouldAddEndPuncttrue
\mciteSetBstMidEndSepPunct{\mcitedefaultmidpunct}
{\mcitedefaultendpunct}{\mcitedefaultseppunct}\relax
\EndOfBibitem
\bibitem[CEPERLEY and ALDER({1980})CEPERLEY, and ALDER]{lda}
CEPERLEY,~D.; ALDER,~B. \emph{{PHYSICAL REVIEW LETTERS}} \textbf{{1980}},
  \emph{{45}}, {566--569}\relax
\mciteBstWouldAddEndPuncttrue
\mciteSetBstMidEndSepPunct{\mcitedefaultmidpunct}
{\mcitedefaultendpunct}{\mcitedefaultseppunct}\relax
\EndOfBibitem
\bibitem[Monkhorst and Pack(1976)Monkhorst, and Pack]{monk}
Monkhorst,~H.~J.; Pack,~J.~D. \emph{PHYSICAL REVIEW B} \textbf{1976},
  \emph{13}, 5188--5192\relax
\mciteBstWouldAddEndPuncttrue
\mciteSetBstMidEndSepPunct{\mcitedefaultmidpunct}
{\mcitedefaultendpunct}{\mcitedefaultseppunct}\relax
\EndOfBibitem
\bibitem[KRESSE and HAFNER({1993})KRESSE, and HAFNER]{vasp1}
KRESSE,~G.; HAFNER,~J. \emph{{PHYSICAL REVIEW B}} \textbf{{1993}}, \emph{{47}},
  {558--561}\relax
\mciteBstWouldAddEndPuncttrue
\mciteSetBstMidEndSepPunct{\mcitedefaultmidpunct}
{\mcitedefaultendpunct}{\mcitedefaultseppunct}\relax
\EndOfBibitem
\bibitem[Kresse and Furthmuller({1996})Kresse, and Furthmuller]{vasp2}
Kresse,~G.; Furthmuller,~J. \emph{{PHYSICAL REVIEW B}} \textbf{{1996}},
  \emph{{54}}, {11169--11186}\relax
\mciteBstWouldAddEndPuncttrue
\mciteSetBstMidEndSepPunct{\mcitedefaultmidpunct}
{\mcitedefaultendpunct}{\mcitedefaultseppunct}\relax
\EndOfBibitem
\bibitem[Shishkin and Kresse({2006})Shishkin, and Kresse]{gw}
Shishkin,~M.; Kresse,~G. \emph{{PHYSICAL REVIEW B}} \textbf{{2006}},
  \emph{{74}}, {035101}\relax
\mciteBstWouldAddEndPuncttrue
\mciteSetBstMidEndSepPunct{\mcitedefaultmidpunct}
{\mcitedefaultendpunct}{\mcitedefaultseppunct}\relax
\EndOfBibitem
\bibitem[YANG et~al.({1991})YANG, SANDOVAL, DIVIGALPITIYA, IRWIN, and
  FRINDT]{Yang}
YANG,~D.; SANDOVAL,~S.; DIVIGALPITIYA,~W.; IRWIN,~J.; FRINDT,~R.
  \emph{{PHYSICAL REVIEW B}} \textbf{{1991}}, \emph{{43}}, {12053--12056}\relax
\mciteBstWouldAddEndPuncttrue
\mciteSetBstMidEndSepPunct{\mcitedefaultmidpunct}
{\mcitedefaultendpunct}{\mcitedefaultseppunct}\relax
\EndOfBibitem
\bibitem[JOENSEN et~al.({1987})JOENSEN, CROZIER, ALBERDING, and
  FRINDT]{Joensen2}
JOENSEN,~P.; CROZIER,~E.; ALBERDING,~N.; FRINDT,~R. \emph{{JOURNAL OF PHYSICS
  C-SOLID STATE PHYSICS}} \textbf{{1987}}, \emph{{20}}, {4043--4053}\relax
\mciteBstWouldAddEndPuncttrue
\mciteSetBstMidEndSepPunct{\mcitedefaultmidpunct}
{\mcitedefaultendpunct}{\mcitedefaultseppunct}\relax
\EndOfBibitem
\bibitem[Mulliken(1955)]{Mulliken}
Mulliken,~R.~S. \emph{The Journal of Chemical Physics} \textbf{1955},
  \emph{23}, 1841--1846\relax
\mciteBstWouldAddEndPuncttrue
\mciteSetBstMidEndSepPunct{\mcitedefaultmidpunct}
{\mcitedefaultendpunct}{\mcitedefaultseppunct}\relax
\EndOfBibitem
\bibitem[Pau()]{Pauli}
It should be kept in mind that there are ambiguities in calculating charge
  transfer. In fact, different methods result in different values of charge
  transfer.\relax
\mciteBstWouldAddEndPunctfalse
\mciteSetBstMidEndSepPunct{\mcitedefaultmidpunct}
{}{\mcitedefaultseppunct}\relax
\EndOfBibitem
\bibitem[Henkelman et~al.({2006})Henkelman, Arnaldsson, and Jonsson]{Bader}
Henkelman,~G.; Arnaldsson,~A.; Jonsson,~H. \emph{{COMPUTATIONAL MATERIALS
  SCIENCE}} \textbf{{2006}}, \emph{{36}}, {354--360}\relax
\mciteBstWouldAddEndPuncttrue
\mciteSetBstMidEndSepPunct{\mcitedefaultmidpunct}
{\mcitedefaultendpunct}{\mcitedefaultseppunct}\relax
\EndOfBibitem
\bibitem[Soler et~al.({2002})Soler, Artacho, Gale, Garcia, Junquera, Ordejon,
  and Sanchez-Portal]{siesta}
Soler,~J.; Artacho,~E.; Gale,~J.; Garcia,~A.; Junquera,~J.; Ordejon,~P.;
  Sanchez-Portal,~D. \emph{{JOURNAL OF PHYSICS-CONDENSED MATTER}}
  \textbf{{2002}}, \emph{{14}}, {2745--2779}\relax
\mciteBstWouldAddEndPuncttrue
\mciteSetBstMidEndSepPunct{\mcitedefaultmidpunct}
{\mcitedefaultendpunct}{\mcitedefaultseppunct}\relax
\EndOfBibitem
\bibitem[KAM and PARKINSON({1982})KAM, and PARKINSON]{3dmos2}
KAM,~K.; PARKINSON,~B. \emph{{JOURNAL OF PHYSICAL CHEMISTRY}} \textbf{{1982}},
  \emph{{86}}, {463--467}\relax
\mciteBstWouldAddEndPuncttrue
\mciteSetBstMidEndSepPunct{\mcitedefaultmidpunct}
{\mcitedefaultendpunct}{\mcitedefaultseppunct}\relax
\EndOfBibitem
\bibitem[Fuchs et~al.(2007)Fuchs, Furthm\"uller, Bechstedt, Shishkin, and
  Kresse]{Kresse}
Fuchs,~F.; Furthm\"uller,~J.; Bechstedt,~F.; Shishkin,~M.; Kresse,~G.
  \emph{Phys. Rev. B} \textbf{2007}, \emph{76}, 115109\relax
\mciteBstWouldAddEndPuncttrue
\mciteSetBstMidEndSepPunct{\mcitedefaultmidpunct}
{\mcitedefaultendpunct}{\mcitedefaultseppunct}\relax
\EndOfBibitem
\bibitem[Heyd et~al.(2003)Heyd, Scuseria, and Ernzerhof]{HSE}
Heyd,~J.; Scuseria,~G.~E.; Ernzerhof,~M. \emph{The Journal of Chemical Physics}
  \textbf{2003}, \emph{118}, 8207--8215\relax
\mciteBstWouldAddEndPuncttrue
\mciteSetBstMidEndSepPunct{\mcitedefaultmidpunct}
{\mcitedefaultendpunct}{\mcitedefaultseppunct}\relax
\EndOfBibitem
\bibitem[\ifmmode~\mbox{\c{S}}\else \c{S}\fi{}ahin
  et~al.(2011)\ifmmode~\mbox{\c{S}}\else \c{S}\fi{}ahin, Topsakal, and
  Ciraci]{CFband}
\ifmmode~\mbox{\c{S}}\else \c{S}\fi{}ahin,~H.; Topsakal,~M.; Ciraci,~S.
  \emph{Phys. Rev. B} \textbf{2011}, \emph{83}, 115432\relax
\mciteBstWouldAddEndPuncttrue
\mciteSetBstMidEndSepPunct{\mcitedefaultmidpunct}
{\mcitedefaultendpunct}{\mcitedefaultseppunct}\relax
\EndOfBibitem
\bibitem[Ataca and Ciraci({2010})Ataca, and Ciraci]{canBN}
Ataca,~C.; Ciraci,~S. \emph{{PHYSICAL REVIEW B}} \textbf{{2010}}, \emph{{82}},
  165402\relax
\mciteBstWouldAddEndPuncttrue
\mciteSetBstMidEndSepPunct{\mcitedefaultmidpunct}
{\mcitedefaultendpunct}{\mcitedefaultseppunct}\relax
\EndOfBibitem
\bibitem[Ataca et~al.({2011})Ataca, Akturk, Sahin, and Ciraci]{canC}
Ataca,~C.; Akturk,~E.; Sahin,~H.; Ciraci,~S. \emph{{JOURNAL OF APPLIED
  PHYSICS}} \textbf{{2011}}, \emph{{109}}, 013704\relax
\mciteBstWouldAddEndPuncttrue
\mciteSetBstMidEndSepPunct{\mcitedefaultmidpunct}
{\mcitedefaultendpunct}{\mcitedefaultseppunct}\relax
\EndOfBibitem
\bibitem[Akturk et~al.({2010})Akturk, Ataca, and Ciraci]{canSi}
Akturk,~E.; Ataca,~C.; Ciraci,~S. \emph{{APPLIED PHYSICS LETTERS}}
  \textbf{{2010}}, \emph{{96}}, 123112\relax
\mciteBstWouldAddEndPuncttrue
\mciteSetBstMidEndSepPunct{\mcitedefaultmidpunct}
{\mcitedefaultendpunct}{\mcitedefaultseppunct}\relax
\EndOfBibitem
\bibitem[Ataca et~al.({2008})Ataca, Akturk, Ciraci, and Ustunel]{canLi}
Ataca,~C.; Akturk,~E.; Ciraci,~S.; Ustunel,~H. \emph{{APPLIED PHYSICS LETTERS}}
  \textbf{{2008}}, \emph{{93}}, 043123\relax
\mciteBstWouldAddEndPuncttrue
\mciteSetBstMidEndSepPunct{\mcitedefaultmidpunct}
{\mcitedefaultendpunct}{\mcitedefaultseppunct}\relax
\EndOfBibitem
\bibitem[Ataca et~al.({2009})Ataca, Akturk, and Ciraci]{canCa}
Ataca,~C.; Akturk,~E.; Ciraci,~S. \emph{{PHYSICAL REVIEW B}} \textbf{{2009}},
  \emph{{79}}, 041406\relax
\mciteBstWouldAddEndPuncttrue
\mciteSetBstMidEndSepPunct{\mcitedefaultmidpunct}
{\mcitedefaultendpunct}{\mcitedefaultseppunct}\relax
\EndOfBibitem
\bibitem[Chan et~al.({2008})Chan, Neaton, and Cohen]{cohen}
Chan,~K.~T.; Neaton,~J.~B.; Cohen,~M.~L. \emph{{PHYSICAL REVIEW B}}
  \textbf{{2008}}, \emph{{77}}, 235430\relax
\mciteBstWouldAddEndPuncttrue
\mciteSetBstMidEndSepPunct{\mcitedefaultmidpunct}
{\mcitedefaultendpunct}{\mcitedefaultseppunct}\relax
\EndOfBibitem
\bibitem[Durgun et~al.({2007})Durgun, Akman, Ataca, and Ciraci]{canNanowire}
Durgun,~E.; Akman,~N.; Ataca,~C.; Ciraci,~S. \emph{{PHYSICAL REVIEW B}}
  \textbf{{2007}}, \emph{{76}}, 245323\relax
\mciteBstWouldAddEndPuncttrue
\mciteSetBstMidEndSepPunct{\mcitedefaultmidpunct}
{\mcitedefaultendpunct}{\mcitedefaultseppunct}\relax
\EndOfBibitem
\bibitem[Novoselov et~al.({2004})Novoselov, Geim, Morozov, Jiang, Zhang,
  Dubonos, Grigorieva, and Firsov]{novos}
Novoselov,~K.; Geim,~A.; Morozov,~S.; Jiang,~D.; Zhang,~Y.; Dubonos,~S.;
  Grigorieva,~I.; Firsov,~A. \emph{{SCIENCE}} \textbf{{2004}}, \emph{{306}},
  {666--669}\relax
\mciteBstWouldAddEndPuncttrue
\mciteSetBstMidEndSepPunct{\mcitedefaultmidpunct}
{\mcitedefaultendpunct}{\mcitedefaultseppunct}\relax
\EndOfBibitem
\bibitem[Sahin et~al.({2009})Sahin, Cahangirov, Topsakal, Bekaroglu, Akturk,
  Senger, and Ciraci]{ansiklopedi}
Sahin,~H.; Cahangirov,~S.; Topsakal,~M.; Bekaroglu,~E.; Akturk,~E.;
  Senger,~R.~T.; Ciraci,~S. \emph{{PHYSICAL REVIEW B}} \textbf{{2009}},
  \emph{{80}}, 155453\relax
\mciteBstWouldAddEndPuncttrue
\mciteSetBstMidEndSepPunct{\mcitedefaultmidpunct}
{\mcitedefaultendpunct}{\mcitedefaultseppunct}\relax
\EndOfBibitem
\bibitem[Cahangirov et~al.({2009})Cahangirov, Topsakal, Akturk, Sahin, and
  Ciraci]{seymur}
Cahangirov,~S.; Topsakal,~M.; Akturk,~E.; Sahin,~H.; Ciraci,~S. \emph{{PHYSICAL
  REVIEW LETTERS}} \textbf{{2009}}, \emph{{102}}, 236804\relax
\mciteBstWouldAddEndPuncttrue
\mciteSetBstMidEndSepPunct{\mcitedefaultmidpunct}
{\mcitedefaultendpunct}{\mcitedefaultseppunct}\relax
\EndOfBibitem
\bibitem[Ataca and Ciraci()Ataca, and Ciraci]{graph+C}
Ataca,~C.; Ciraci,~S. \emph{{arXiv:1012.1185v3}} \relax
\mciteBstWouldAddEndPunctfalse
\mciteSetBstMidEndSepPunct{\mcitedefaultmidpunct}
{}{\mcitedefaultseppunct}\relax
\EndOfBibitem
\bibitem[Esquinazi et~al.({2003})Esquinazi, Spemann, Hohne, Setzer, Han, and
  Butz]{esquinazi}
Esquinazi,~P.; Spemann,~D.; Hohne,~R.; Setzer,~A.; Han,~K.; Butz,~T.
  \emph{{PHYSICAL REVIEW LETTERS}} \textbf{{2003}}, \emph{{91}}, {227201}\relax
\mciteBstWouldAddEndPuncttrue
\mciteSetBstMidEndSepPunct{\mcitedefaultmidpunct}
{\mcitedefaultendpunct}{\mcitedefaultseppunct}\relax
\EndOfBibitem
\bibitem[Hashimoto et~al.({2004})Hashimoto, Suenaga, Gloter, Urita, and
  Iijima]{Iijima}
Hashimoto,~A.; Suenaga,~K.; Gloter,~A.; Urita,~K.; Iijima,~S. \emph{{NATURE}}
  \textbf{{2004}}, \emph{{430}}, {870--873}\relax
\mciteBstWouldAddEndPuncttrue
\mciteSetBstMidEndSepPunct{\mcitedefaultmidpunct}
{\mcitedefaultendpunct}{\mcitedefaultseppunct}\relax
\EndOfBibitem
\bibitem[Yazyev and Helm({2007})Yazyev, and Helm]{yazyev}
Yazyev,~O.~V.; Helm,~L. \emph{{PHYSICAL REVIEW B}} \textbf{{2007}},
  \emph{{75}}, {125408}\relax
\mciteBstWouldAddEndPuncttrue
\mciteSetBstMidEndSepPunct{\mcitedefaultmidpunct}
{\mcitedefaultendpunct}{\mcitedefaultseppunct}\relax
\EndOfBibitem
\bibitem[Vozmediano et~al.({2005})Vozmediano, Lopez-Sancho, Stauber, and
  Guinea]{guinea}
Vozmediano,~M.; Lopez-Sancho,~M.; Stauber,~T.; Guinea,~F. \emph{{PHYSICAL
  REVIEW B}} \textbf{{2005}}, \emph{{72}}, {155121}\relax
\mciteBstWouldAddEndPuncttrue
\mciteSetBstMidEndSepPunct{\mcitedefaultmidpunct}
{\mcitedefaultendpunct}{\mcitedefaultseppunct}\relax
\EndOfBibitem
\bibitem[Brey et~al.({2007})Brey, Fertig, and Das~Sarma]{brey2}
Brey,~L.; Fertig,~H.~A.; Das~Sarma,~S. \emph{{PHYSICAL REVIEW LETTERS}}
  \textbf{{2007}}, \emph{{99}}, {116802}\relax
\mciteBstWouldAddEndPuncttrue
\mciteSetBstMidEndSepPunct{\mcitedefaultmidpunct}
{\mcitedefaultendpunct}{\mcitedefaultseppunct}\relax
\EndOfBibitem
\bibitem[Palacios et~al.({2008})Palacios, Fernandez-Rossier, and Brey]{brey1}
Palacios,~J.~J.; Fernandez-Rossier,~J.; Brey,~L. \emph{{PHYSICAL REVIEW B}}
  \textbf{{2008}}, \emph{{77}}, {195428}\relax
\mciteBstWouldAddEndPuncttrue
\mciteSetBstMidEndSepPunct{\mcitedefaultmidpunct}
{\mcitedefaultendpunct}{\mcitedefaultseppunct}\relax
\EndOfBibitem
\bibitem[Topsakal et~al.({2008})Topsakal, Akturk, Sevincli, and Ciraci]{delik}
Topsakal,~M.; Akturk,~E.; Sevincli,~H.; Ciraci,~S. \emph{{PHYSICAL REVIEW B}}
  \textbf{{2008}}, \emph{{78}}, {235435}\relax
\mciteBstWouldAddEndPuncttrue
\mciteSetBstMidEndSepPunct{\mcitedefaultmidpunct}
{\mcitedefaultendpunct}{\mcitedefaultseppunct}\relax
\EndOfBibitem
\bibitem[Sahin et~al.({2009})Sahin, Ataca, and Ciraci]{sahingraphane}
Sahin,~H.; Ataca,~C.; Ciraci,~S. \emph{{APPLIED PHYSICS LETTERS}}
  \textbf{{2009}}, \emph{{95}}, {222510}\relax
\mciteBstWouldAddEndPuncttrue
\mciteSetBstMidEndSepPunct{\mcitedefaultmidpunct}
{\mcitedefaultendpunct}{\mcitedefaultseppunct}\relax
\EndOfBibitem
\bibitem[Sahin et~al.({2010})Sahin, Ataca, and Ciraci]{sahinprb}
Sahin,~H.; Ataca,~C.; Ciraci,~S. \emph{{PHYSICAL REVIEW B}} \textbf{{2010}},
  \emph{{81}}, {205417}\relax
\mciteBstWouldAddEndPuncttrue
\mciteSetBstMidEndSepPunct{\mcitedefaultmidpunct}
{\mcitedefaultendpunct}{\mcitedefaultseppunct}\relax
\EndOfBibitem
\bibitem[LIEB({1989})]{lieb}
LIEB,~E. \emph{{PHYSICAL REVIEW LETTERS}} \textbf{{1989}}, \emph{{62}},
  {1201--1204}\relax
\mciteBstWouldAddEndPuncttrue
\mciteSetBstMidEndSepPunct{\mcitedefaultmidpunct}
{\mcitedefaultendpunct}{\mcitedefaultseppunct}\relax
\EndOfBibitem
\bibitem[Bai et~al.({2010})Bai, Zhong, Jiang, Huang, and Duan]{bai}
Bai,~J.; Zhong,~X.; Jiang,~S.; Huang,~Y.; Duan,~X. \emph{{NATURE
  NANOTECHNOLOGY}} \textbf{{2010}}, \emph{{5}}, {190--194}\relax
\mciteBstWouldAddEndPuncttrue
\mciteSetBstMidEndSepPunct{\mcitedefaultmidpunct}
{\mcitedefaultendpunct}{\mcitedefaultseppunct}\relax
\EndOfBibitem
\bibitem[Lahiri et~al.({2010})Lahiri, Lin, Bozkurt, Oleynik, and
  Batzill]{lahiri}
Lahiri,~J.; Lin,~Y.; Bozkurt,~P.; Oleynik,~I.~I.; Batzill,~M. \emph{{NATURE
  NANOTECHNOLOGY}} \textbf{{2010}}, \emph{{5}}, {326--329}\relax
\mciteBstWouldAddEndPuncttrue
\mciteSetBstMidEndSepPunct{\mcitedefaultmidpunct}
{\mcitedefaultendpunct}{\mcitedefaultseppunct}\relax
\EndOfBibitem
\bibitem[Balog et~al.({2010})Balog, Jorgensen, Nilsson, Andersen, Rienks,
  Bianchi, Fanetti, Laegsgaard, Baraldi, Lizzit, Sljivancanin, Besenbacher,
  Hammer, Pedersen, Hofmann, and Hornekaer]{balog}
Balog,~R. et~al.  \emph{{NATURE MATERIALS}} \textbf{{2010}}, \emph{{9}},
  {315--319}\relax
\mciteBstWouldAddEndPuncttrue
\mciteSetBstMidEndSepPunct{\mcitedefaultmidpunct}
{\mcitedefaultendpunct}{\mcitedefaultseppunct}\relax
\EndOfBibitem
\end{mcitethebibliography}

\end{document}